\documentclass[useAMS,usenatbib]{mn2e}

\usepackage{url}
\usepackage{graphicx}
\usepackage{longtable}
\usepackage{lscape}
\usepackage{comment}
\usepackage{amssymb,amsmath}
\usepackage{color}

\newcommand{\revone}{}
\newcommand{\revtwo}{}

\newcommand{\mstar}{\mbox{$M_{\rm star}$}}
\newcommand{\submm}{submm}
\newcommand{\Submm}{Submm}
\newcommand{\mm}{millimetre}
\newcommand{\smgs}{{\submm} galaxies}
\newcommand{\smg}{{\submm} galaxy}
\newcommand{\Smgs}{{\Submm} galaxies}

\newcommand{\msun}{\mbox{M$_\odot$}}
\newcommand{\lsun}{\mbox{L$_\odot$}}
\newcommand{\msunyr}{\mbox{\msun\ yr$^{-1}$}}
\newcommand{\pergyr}{\mbox{Gyr$^{-1}$}}

\def\aap{A\hbox{\rm \&}A} 
\def\aaps{A\hbox{\rm \&}A Suppl.}   
  \def\apj{ApJ} \def\apjl{ApJ}
 \def\apjs{ApJS} 
\def\araa{ARA\hbox{\rm \&}A}

\def\mnras{MNRAS} \def\nat{Nat} \def\pasj{PASJ}

\def\physrep{{Phys.~Rep.}}   

\title[S2CLS: submm galaxies in 2\,deg$^2$ 850-$\mu$m imaging]{%
\vspace{-0.5cm}%
The SCUBA-2 Cosmology Legacy Survey: the nature of bright submm galaxies 
from 2\,deg$^2$  of 850-$\mu$m imaging
\vspace{-0.5em}%
}
\author[M.~J.~Micha{\l}owski et al.]%
{Micha{\l}~J.~Micha{\l}owski$^{1}$\thanks{E-mail: mj.michalowski@gmail.com},
J.~S.~Dunlop$^{1}$,
M.~P.~Koprowski$^{1,2}$,
M.~Cirasuolo$^{3,4,1}$,
\newauthor
J.~E.~Geach$^{2}$,
R.~A.~A.~Bowler$^{1,5}$,
A.~Mortlock$^{1}$, 
K.~I.~Caputi$^{6}$,
I.~Aretxaga$^{7}$,
\newauthor
V.~Arumugam$^{3,1}$,
Chian-Chou~Chen$^{8,3}$,
R.~J.~McLure$^{1}$,
M.~Birkinshaw$^{9,10}$,
\newauthor
N.~Bourne$^1$,
D.~Farrah$^{11}$,
E.~Ibar$^{12}$,
P.~van der Werf$^{13}$,
M.~Zemcov$^{14}$
\\
$^{1}$SUPA\thanks{Scottish Universities Physics Alliance}, Institute for Astronomy, University of Edinburgh, Royal Observatory, Blackford Hill, Edinburgh, EH9 3HJ, UK\\
$^2$Center for Astrophysics Research, Science and Technology Research Institute, University of Hertfordshire, Hatfield AL10 9AB, UK\\
$^3$European Southern Observatory, Karl Schwartzchild Strasse 2, D-85748 Garching, Germany\\
$^4$UK Astronomy Technology Centre, Royal Observatory, Edinburgh EH9 3HJ, UK\\
$^5$Astrophysics, The Denys Wilkinson Building, University of Oxford, Keble Road, Oxford OX1 3RH, UK\\
$^6$Kapteyn Astronomical Institute, University of Groningen, PO Box 800, NL-9700 AV Groningen, The Netherlands\\
$^7$Instituto Nacional de Astrof\'{\i}sica, \'Optica y Electr\'onica (INAOE), Aptdo. Postal 51 y 216, 72000 Puebla, Pue., Mexico  \\
$^8$Centre for Extragalactic Astronomy, Department of Physics, Durham University, South Road, Durham DH1 3LE, UK\\
$^9$HH Wills Physics Laboratory, University of Bristol, Tyndall Avenue, Bristol BS8 1TL, UK\\
$^{10}$Harvard-Smithsonian Center for Astrophysics, 60 Garden Street, Cambridge, MA 02138, USA\\
$^{11}$Department of Physics, Virginia Tech, Blacksburg, VA 24061, USA\\
$^{12}$Instituto de F\'isica y Astronom\'ia, Universidad de Valpara\'iso, Avda.\ Gran Breta\~na 1111, Valparaiso, Chile\\
$^{13}$Leiden Observatory, Leiden University, PO Box 9513, NL-2300 RA Leiden, the Netherlands\\
$^{14}$Center for Detectors, School of Physics and Astronomy, Rochester Institute of Technology, Rochester, NY 14623, USA
\vspace{-0.5cm}
}
\begin{document}

\date{Accepted 2017 April 5. Received 2017 March 30; in original form 2016 October 7.}

\pagerange{\pageref{firstpage}--\pageref{lastpage}} \pubyear{2016}

\maketitle

\label{firstpage}

\vspace*{-0.5cm}

\begin{abstract}
We present physical properties [redshifts ($z$), star formation rates (SFRs) and stellar masses ($\mstar$)] of bright ($S_{850} \ge 4$\,mJy) {\submm} galaxies  in the $\simeq 2\,\mbox{deg}^2$ COSMOS and UDS fields selected with SCUBA-2/JCMT.
{\revone We complete the galaxy identification process for all ($\simeq 2\,000$) S/N~$\ge$~3.5 850-$\mu$m sources, but focus our scientific analysis on a high-quality subsample of 651 S/N~$\ge$~4 sources with complete multiwavelength coverage including 1.1-mm imaging. }
We check the reliability of our identifications, and the robustness of the SCUBA-2  fluxes by revisiting the 
recent ALMA follow-up of {\revone 29 sources in our sample. }
Considering $>4\,$mJy ALMA sources, 
our identification method has a completeness
of $\simeq 86$ per cent with a reliability of $\simeq 92$ per cent, and 
only $\simeq 15$--$20$ per cent of sources
are significantly affected by multiplicity (when a secondary component contributes $>1/3$ of the primary source flux). 
The impact of source blending on
the 850-$\mu$m source counts as determined with SCUBA-2 is modest; scaling the single-dish
fluxes by $\simeq 0.9$ reproduces the ALMA source counts. 
{\revone For our final SCUBA-2 sample,} we find median $z = 2.40^{+0.10}_{-0.04}$, $\mbox{SFR} = 287\pm6\,\msunyr$
and $\log(\mstar/\msun) = 11.12\pm0.02$ {\revtwo (the latter for 349/651 sources with optical identifications)}. 
These properties clearly locate
bright {\submm} galaxies on the high-mass end of the `main sequence' of star-forming galaxies
out to $z \simeq 6$,  
suggesting that major mergers are not a dominant driver of the high-redshift {\submm}-selected population. 
Their number densities 
are also consistent with the evolving
galaxy stellar mass function. Hence, 
the {\submm} galaxy population is 
as expected, albeit reproducing the evolution of the main sequence of star-forming galaxies 
remains a challenge for theoretical models/simulations.

\end{abstract}


\newpage

\section{Introduction}

Since their discovery almost twenty years ago \citep{smail97,hughes98,barger98}, the nature of galaxies selected at submillimetre ({\submm})
wavelengths ({\smgs}), and their role in galaxy evolution, has been the subject of extensive study
(see  \citealt*{casey14} and \citealt{blain02} for reviews). Of particular importance is the determination of the mechanism
that drives the huge star formation rates (SFRs, and hence huge far-infrared luminosities) of these galaxies, in order to better understand their formation
and subsequent evolution.

This can be studied using various different diagnostics, including the location of galaxies on the stellar mass ($\mstar$) versus SFR plane. At a given redshift, normal star-forming galaxies form a so-called main sequence on this plane
(with near constant specific star formation rate, $\mbox{sSFR}\equiv\mbox{SFR}/\mstar$), whereas `starbursts' are offset
towards higher sSFRs by a factor of $>2$--$4$ \citep{daddi07,noeske07,gonzalez10,elbaz11,speagle14}. Hence, the location of
{\smgs} with respect to the main sequence may tell us whether they are predominantly triggered by mergers, or alternatively
are fed by (relatively steady) cold gas infall \citep[the two options proposed by theoretical arguments;][]{swinbank08,dave10,narayanan10,narayanan15,ricciardelli10,gonzalez11,
hayward11b,hayward11,hayward12,cowley15}. This is because a major merger is a short-lived phenomenon, resulting in a substantial but
temporary boost in SFR, potentially pushing a galaxy significantly above the main sequence (e.g.~\citealt{hung13}; cf.~\citealt{forsterschreiber09}).
{\revone Recent simulations show that high-redshift gas-rich mergers result in the SFR enhancement by a factor of $\sim2$--$5$ \citep[][their figs 5--7]{fensch17}, so if {\smgs} are predominantly powered by major mergers, then they should by offset from the main sequence by this factor.}

There is still some debate over whether {\smgs} lie above the main sequence, or simply form its high-mass end.
{\revone This debate is not primarily concerned with the form of the main sequence, as most studies agree that, at high redshifts, the main sequence continues to extend to high stellar masses with $\mbox{SFR}\propto M_*^x$, where $x$ is in the range $0.75$--$1.0$ \citep{karim11,speagle14,renzini15,schreiber15,koprowski16,dunlop17} with no evidence of any break {\revtwo as has been suggested at lower redshifts \citep{oliver10b, whitaker14,ilbert15,lee15,tomczak16}, or from galaxy surveys based purely on optical data \citep{kochiashvili15,tasca15}}. Based on morphological decomposition at low redshifts this break was shown to disappear when only disc (not bulge) stellar mass was used \citep{abramson14}.}
Low stellar mass estimates for {\smgs} lead to high sSFRs, above the main sequence \citep{hainline11,wardlow11,magnelli12,casey13}, whereas
higher derived stellar masses place {\smgs} on the main sequence \citep{michalowski10smg,michalowski12,michalowski14mass,yun12,
johnson13,koprowski14,koprowski16}. In \citet{michalowski12mass} we showed that this discrepancy results largely from different
assumptions concerning the parametrization of star formation histories in the spectral energy distribution (SED) modelling.
{\revone In particular,  two-component star formation histories result in higher stellar masses. Such a choice assumes that galaxies before the beginning of the {\smg} phase (either a peak of the gas accretion or a merger) have already built up a substantial fraction of  their current stellar mass. }
In \citet{michalowski14mass} we showed that two-component star formation histories (resulting in higher stellar masses)
provide the most accurate stellar masses for a sample of simulated {\smgs}, which have properties that agree well with
many properties of real {\smgs} \citep[][and references therein]{michalowski14mass}. 

Hence, our studies of medium-size samples
of around a hundred {\smgs} resulted in the conclusion that they form the high-mass end of the main sequence, at least at
$z\lesssim3$--$4$ \citep{michalowski12,koprowski14,koprowski16}.
A similar conclusion has been drawn from recent hydrodynamical simulations showing that all observational properties of
{\smgs} can be explained by non-merging massive galaxies that sustain high SFRs for around 1\,Gyr, and do not leave the
main sequence during that time (\citealt{narayanan15}; see also \citealt{dave10,hayward11b,shimizu12}). 
{\revone However, other simulations predict that a significant fraction of {\smgs} are powered by violent starbursts resulting from mergers \citep{baugh05,narayanan10,hayward13}.}
Further observational studies based on larger samples of {\smgs} are required
to clarify this issue.

In addition, rather little is known about the very high-redshift ($z>4$) tail of the submm galaxy population, because to date
only a handful of submm sources have been confirmed at these extreme redshifts \citep{coppin09,capak08,capak11,schinnerer08,
daddi09,daddi09b,knudsen08b,knudsen09,riechers10,riechers13,cox11,smolcic11,combes12,
walter12b,dowell14,watson15}. The discovery and study of such sources is difficult for several reasons. First,
these very high-redshift sources are intrinsically rare, so very few of them are likely to be discovered in submm surveys covering only a
few hundred square arcmin (as typically achieved at 850$\mu$m prior to SCUBA-2). Secondly, the combined effects of extreme dust-obscuration
and redshift mean that optical and radio counterparts can be extremely faint \citep[e.g.][]{walter12b}, and hence redshift
information hard to secure. Moreover, the 
determination of redshifts at submm/mm wavelengths from carbon monoxide (CO) lines
currently remains very time consuming for all but the brightest objects \citep{weiss09,weiss13,vieira13}, and hence is not practical
for large samples. These difficulties, and the resulting small samples of confirmed 
high-redshift {\smgs} have also hampered
the proper statistical investigation of suggestions that the brightest submm sources are preferentially found at the highest redshifts
\citep{ivison02,pope05,michalowski12,koprowski14,simpson14}.

Our understanding of both the relation of {\smgs} with respect to the main sequence, and the prevalence and nature of the most
extreme redshift submm sources can both be improved by the larger area {\submm} surveys now being provided by SCUBA-2. Hence, here we use the largest deep survey at 850 $\mu$m undertaken to date,
the SCUBA-2 Cosmology Legacy Survey (CLS). This survey is described, and the 850$\mu$m catalogues are presented in \citet{geach17}.
The results from smaller, deeper sub-fields within the CLS have already been presented in \citet{geach13},  \citet{roseboom13},  \citet{koprowski16} and \citet{zavala17},
while multiwavelength identifications (IDs) for the sources in the $\sim1\,\mbox{deg}^2$ UDS field have been provided by \citet{chen16}.

Here we build on this work by attempting to determine the identifications, redshifts and physical properties of a statistically significant, well-defined sample of
around 2000 {\smgs} detected in the full $\sim2\,\mbox{deg}^2$ of 850-$\mu$m imaging provided by the S2CLS across the UDS and COSMOS fields.
A key objective of this study is to assemble a substantial but well-defined subsample of submm sources with complete redshift information, in order to
better define the high-redshift tail of the population, and to clarify the extent to which submm galaxies can indeed be naturally explained
by the high-mass end of the evolving main sequence of star-forming galaxies.

This paper is structured as follows. In Section~\ref{sec:data} we summarise the submm imaging, and describe the supporting higher-resolution
multiwavelength data (optical/near-IR/mid-IR/radio) that we utilise to establish the positions of the galaxy counterparts to the submm sources
in the two survey fields. In Section~\ref{sec:id}, we describe the methods used to identify potential galaxy counterparts,
and to assess their statistical  significance/robustness. In Section 4 we then pause to revisit the results of existing ALMA follow-up of 29
of the sources in our sample, both to assess the robustness and completeness of our identification process, and to assess the impact of
source multiplicity/blending on the reliability of the 850-$\mu$m source counts. In Section 5 we discuss and present the long-wavelength imaging available
in our survey fields; such information is crucial for the estimation of redshifts for sources that lack optical/near-IR counterparts, and for the
estimation of SFRs, and leads us to define a subset of 651 sources with
the information required for an unbiased investigation of their physical properties (i.e. with $\ge4\sigma$ detections at 850 $\mu$m, and sufficient
multiwavelength data to yield complete/unbiased redshift information).
The photometric redshifts, and source number density as a function of redshift are derived in Section~6, while
SFRs and stellar masses are presented in Section~7. We discuss the implications of our results in Section~8, and close with our conclusions in Section~9.
We use a cosmological model with $H_0=70$\,km\,s$^{-1}$\,Mpc$^{-1}$,  $\Omega_\Lambda=0.7$ and $\Omega_m=0.3$, and 
give all magnitudes in the AB system \citep{magab}.

\section{Data}
\label{sec:data}

\begin{table}
\caption{The $3\sigma$ depths of the multifrequency data used in the COSMOS and UDS fields.}
\label{tab:data}
\begin{center}
\begin{tabular}{lccl}
\hline\hline
Filter			&COSMOS 		& UDS	& Unit 		\\
\hline
$u$			& 27.1	& $\cdots$& AB mag\\	
$B$			& $\cdots$& 27.8	& AB mag		\\
$V$			& $\cdots$& 27.4	& AB mag		\\
$g$			& 27.2	& $\cdots$& AB mag\\	
$r$			& 26.7	& $\cdots$& AB mag\\	
$R$			& $\cdots$& 27.1	& AB mag		\\
$i$			& 26.4	& 27.0	& AB mag		\\
$z'$			& 25.3	& 26.3	& AB mag		\\
$Y$			& 25.0/25.6$^a$	& 25.1	& AB mag		\\
$J$			& 24.9/25.2$^a$	& 25.6	& AB mag		\\
$H$			& 24.5/24.8$^a$	& 25.1	& AB mag		\\
$K_s$		& 24.0/24.9$^a$	& 25.2	& AB mag		\\
$3.6\,\mu$m	& 0.17	& 0.18	& $\mu$Jy	\\	
$4.5\,\mu$m	& 0.20	& 0.22	& $\mu$Jy	\\	
$5.6\,\mu$m	& 6.8		& 19		& $\mu$Jy	\\	
$8.0\,\mu$m	& 8.8		& 12		& $\mu$Jy	\\	
$24\,\mu$m	& 40		& 30		& $\mu$Jy	\\
$100\,\mu$m	& 4.6		& 6.7		& mJy	\\
$160\,\mu$m	& 8.8   	& 12.8	& mJy	\\
$250\,\mu$m	& 18		& 19		& mJy	\\
$350\,\mu$m	& 19		& 20		& mJy	\\
$500\,\mu$m	& 21		& 22		& mJy	\\
$850\,\mu$m	& 4.3		& 2.7		& mJy	\\
$1.1$ mm		& 3.8 	& 3.0--5.1	& mJy		\\
$1.4$ GHz	& 36		& 27		& $\mu$Jy	\\
\hline 
\end{tabular}

$^a$ The two alternative values correspond to the shallower and deeper strips of the UltraVISTA near-IR imaging. 
\end{center}
\end{table}



\subsection{Submm}

We used the $850\,\micron$ data obtained with the James Clerk Maxwell Telescope (JCMT) equipped with the
Submillimetre Common-User Bolometer Array 2 \citep[SCUBA-2;][]{scuba2} within the Cosmology Legacy Survey (CLS; \citealt{geach17}). 
The SCUBA-2 data were reduced with the {\sc Smurf}\footnote{\url{www.starlink.ac.uk/docs/sun258.htx/sun258.html}} package
V1.4.0 \citep{chapin13} with the flux calibration factor (FCF) of 537\,Jy\,pW$^{-1}$\,beam$^{-1}$ 
\citep{dempsey13}. The full width at half-maximum (FWHM) of the resulting $850\,\micron$ map is 14.6\,arcsec.
 
For this study we have used the `wide' SCUBA-2 $850\,\mu$m maps of the COSMOS
($1.22$\,deg$^2$ reaching $\simeq 1.4\,$mJy rms) and UDS ($0.96$\,deg$^2$ reaching $\simeq 0.9\,$mJy rms) fields.
{\revone They were selected because they are the two largest CLS fields corresponding to $\sim70$\% of the total survey area, and because in most of the other (smaller) fields the auxiliary data are shallower, making it more difficult to constrain physical properties of {\smgs}.} 
The source catalogue is presented in \citet{geach17}, who extracted the sources by searching for
peaks in  the beam-convolved map with a signal-to-noise ratio $\geq 3.5\sigma$.
This process resulted in 726 and 1088 sources in the COSMOS and UDS fields, respectively.
The source S2CLSJ021830-053130 with an $850\mu$m flux of $\sim50$\,mJy is the lensed candidate discussed
by  \citet{ikarashi11}.

\subsection{Radio and mid-infrared}
 
The Karl G. Jansky Very Large Array (VLA) $1.4$\,GHz radio data were taken from 
\citet{schinnerer07,schinnerer10} for the COSMOS field, and from 
\citet[][]{ivison05,ivison07} and Arumugam et al.~(in preparation) for the UDS field. 
The catalogues include sources for which $\ge3\sigma$ detections were obtained.

The mid-infrared (mid-IR) {\it Spitzer} \citep{spitzer,irac,mips} data 
are from the {\it Spitzer} Extended Deep Survey \citep[SEDS;][]{ashby13},
the {\it Spitzer} Large Area Survey with Hyper-Suprime-Cam (SPLASH, PI: P.~Capak),
the S-COSMOS project \citep{sanders07,lefloch09}
and the {\it Spitzer} Public Legacy Survey of the UKIDSS  Ultra Deep Survey 
(SpUDS; PI: J.~Dunlop)\footnote{\url{ssc.spitzer.caltech.edu/spitzermission/}
\url{observingprograms/legacy/spuds/}} described in \citet{caputi11}.
To obtain the $3.6$ and $4.5\,\micron$ photometry we used the de-confusion code
T-PHOT\footnote{\url{www.astrodeep.eu/t-phot/}} \citep{tphot}. This utilises prior information
on the positions and morphologies of objects from a high-resolution image (HRI; in this case
the $K$-band or $K_s$-band images)
to construct a model of a given low-resolution image (LRI; in this case the {\it Spitzer} imaging)
while solving for the fluxes of these objects. 

The $3\sigma$ depths of the VLA radio and {\it Spitzer} mid-IR imaging
in both fields are summarised in Table\,1.

\subsection{Optical and near-infrared}
\label{sec:dataopt}

The optical data in both fields were obtained with Subaru/SuprimeCam \citep{suprimecam},
as described in \citet{dye06} and \citet{furusawa08}, and from the Canada-France-Hawaii Telescope
Legacy Survey (CFHTLS), as described in \citet{bowler12}. The deep $z'$-band images are
described in \citet{bowler12} and \citet{furusawa16}. 
The near-infrared (near-IR) data in the COSMOS field was obtained from Data Release 2 of
the UltraVISTA survey \citep{mccracken12,bowler14}, while in the UDS field the near-IR
data were provided by Data Release 10 of the UKIRT 
Infrared Deep Sky Survey  \citep[UKIDSS;][]{lawrence07,cirasuolo10,fontana14}.

In both fields the optical and near-IR fluxes were measured in 3-arcsec diameter apertures, and the
resulting $3\sigma$ depths of this aperture photometry are summarised in Table~\ref{tab:data}.

{\revone Finally, we used a list of spectroscopic redshifts from 
3D-HST \citep{brammer12,skelton14,momcheva16},
VIMOS Ultra-Deep Survey \citep[VUDS;][]{lefevre15,tasca17},
zCOSMOS \citep{lilly07,lilly09},
MOSFIRE Deep Evolution Field \citep[MOSDEF;][]{kriek15},
PRIsm MUlti-object Survey \citep[PRIMUS;][]{coil11}
and from \citet{trump09,trump11}
in the COSMOS field and
UDSz \citep[][Almaini et al., in preparation]{mclure13,bradshaw13} and
VIMOS Public Extragalactic Redshift Survey \citep[VIPERS;][]{guzzo14} in the UDS field.
}

\section[]{Galaxy Identifications}
\label{sec:id}

\begin{table*}
\caption{Galaxy counterpart identification statistics for SCUBA-2 sources in the UDS and COSMOS fields, detailing success
rates for both robust and tentative IDs, split by wavelength, and also tabulated for three different significance cuts in
the original 850-$\mu$m sample.
}
\label{tab:succ}
\begin{tiny}
\begin{tabular}{lrccccccccccccrc}
\hline\hline
Field                  & \multicolumn{1}{c}{N}   & rob. ID & tent. ID &No ID & \multicolumn{1}{c}{N$_{1.4}$} & rob$_{1.4}$ & tent$_{1.4}$ & \multicolumn{1}{c}{N$_{24}$} & rob$_{24}$ & tent$_{24}$ & \multicolumn{1}{c}{N$_{8}$} & rob$_{8}$ & tent$_{8}$ & \multicolumn{1}{c}{N$_{\rm opt}$} & $z_{\rm opt}$ \\
                           &                                           & \# (\%) & \# (\%) & \# (\%) & & \# (\%) & \# (\%) & & \# (\%) & \# (\%) & & \# (\%) & \# (\%) & & \# (\%) \\
(1) &  \multicolumn{1}{c}{(2)} & (3) & (4) & (5) & (6) & (7) & (8) & (9) & (10) & (11) & \multicolumn{1}{c}{(12)} & (13) &(14) & (15) &(16) \\
\hline
\multicolumn{16}{l}{\bf S/N$_{850}$$\ge$3.5}\\
COSMOS & 726 & 376 (52) & 96 (13) & 254 (35) & 694 & 181 (26) & 0 (0) & 700 & 290 (41) & 111 (16) & 719 & 189 (26) & 137 (19) & 448 & 310 (69) \\
UDS & 1088 & 546 (50) & 178 (16) & 364 (33) & 1084 & 307 (28) & 25 (2) & 963 & 415 (43) & 172 (18) & 951 & 261 (27) & 191 (20) & 968 & 616 (64) \\
Both & 1814 & 922 (51) & 274 (15) & 618 (34) & 1778 & 488 (27) & 25 (1) & 1663 & 705 (42) & 283 (17) & 1670 & 450 (27) & 328 (20) & 1416 & 926 (65) \\

\hline
\multicolumn{16}{l}{\bf S/N$_{850}$$\ge$4}\\
COSMOS & 405 & 252 (62) & 51 (13) & 102 (25) & 393 & 133 (34) & 0 (0) & 392 & 194 (49) & 62 (16) & 401 & 132 (33) & 90 (22) & 265 & 208 (78) \\
UDS & 716 & 397 (55) & 115 (16) & 204 (28) & 714 & 231 (32) & 17 (2) & 635 & 302 (48) & 117 (18) & 621 & 192 (31) & 137 (22) & 643 & 435 (68) \\
Both & 1121 & 649 (58) & 166 (15) & 306 (27) & 1107 & 364 (33) & 17 (2) & 1027 & 496 (48) & 179 (17) & 1022 & 324 (32) & 227 (22) & 908 & 643 (71) \\

\hline 
\multicolumn{16}{l}{\bf S/N$_{850}$$\ge$5}\\
COSMOS & 185 & 138 (75) & 18 (10) & 29 (16) & 182 & 81 (45) & 0 (0) & 181 & 113 (62) & 19 (10) & 183 & 86 (47) & 51 (28) & 124 & 106 (85) \\
UDS & 333 & 209 (63) & 41 (12) & 83 (25) & 332 & 144 (43) & 6 (2) & 306 & 149 (49) & 65 (21) & 299 & 105 (35) & 72 (24) & 309 & 218 (71) \\
Both & 518 & 347 (67) & 59 (11) & 112 (22) & 514 & 225 (44) & 6 (1) & 487 & 262 (54) & 84 (17) & 482 & 191 (40) & 123 (26) & 433 & 324 (75) \\

\hline 
\end{tabular}
\end{tiny}
{
(1) field name; (2) the total number of SCUBA-2 sources, (3) the number of sources with IDs having at least one robust
association with $p\le0.05$ at radio, $24\,\mu$m, or $8.0\,\mu$m; (4) the number of sources with IDs having at least one tentative  counterpart with $0.05<p<0.1$; (5)  the number of sources with no IDs; 
(6) the number of SCUBA-2 sources covered by the radio map (for which radio IDs can in principle be obtained);
(7) and (8) the number of robust and tentative $1.4\,$GHz IDs;
(9) the number of SCUBA-2 sources covered by the $24\,\micron$ map (for which $24\,\micron$ IDs can in principle be obtained);
(10) and (11) the number of robust and tentative $24\,\mu$m IDs; 
(12) the number of SCUBA-2 sources covered by the $8.0\,\micron$ map (for which $8.0\,\micron$ IDs can in principle be obtained);
(13) and (14) the number of robust and tentative $8.0\,\mu$m IDs; 
(15) the number of SCUBA-2 sources covered by the optical map (for which optical redshift can in principle be derived);
(16) the number of SCUBA-2 sources with the best ID having an optical redshift.
In the parentheses the percentage of IDs are shown.
}
\end{table*}

\begin{figure*}
\begin{tabular}{cc}
\includegraphics[width=0.45\textwidth]{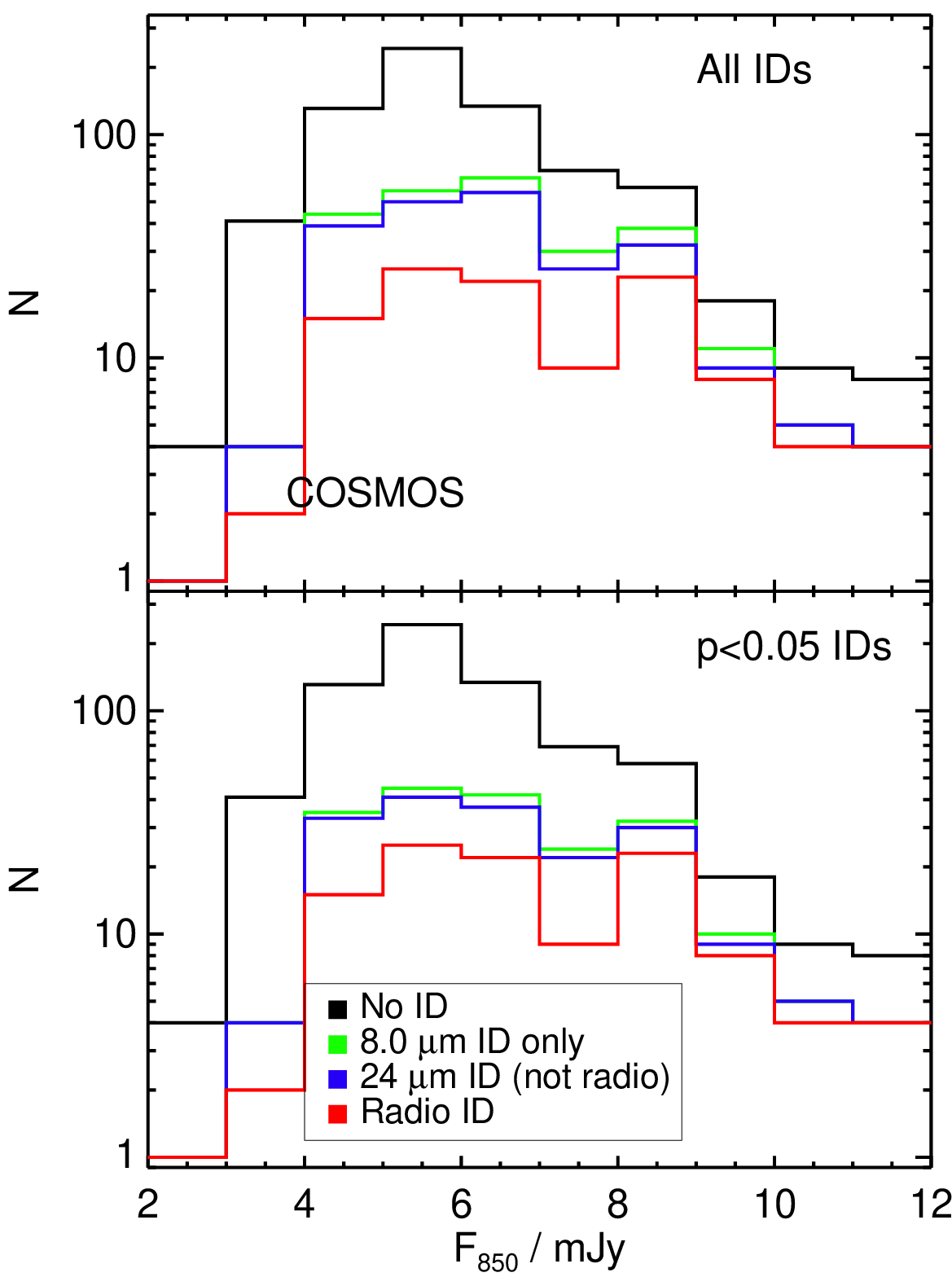} & \includegraphics[width=0.45\textwidth]{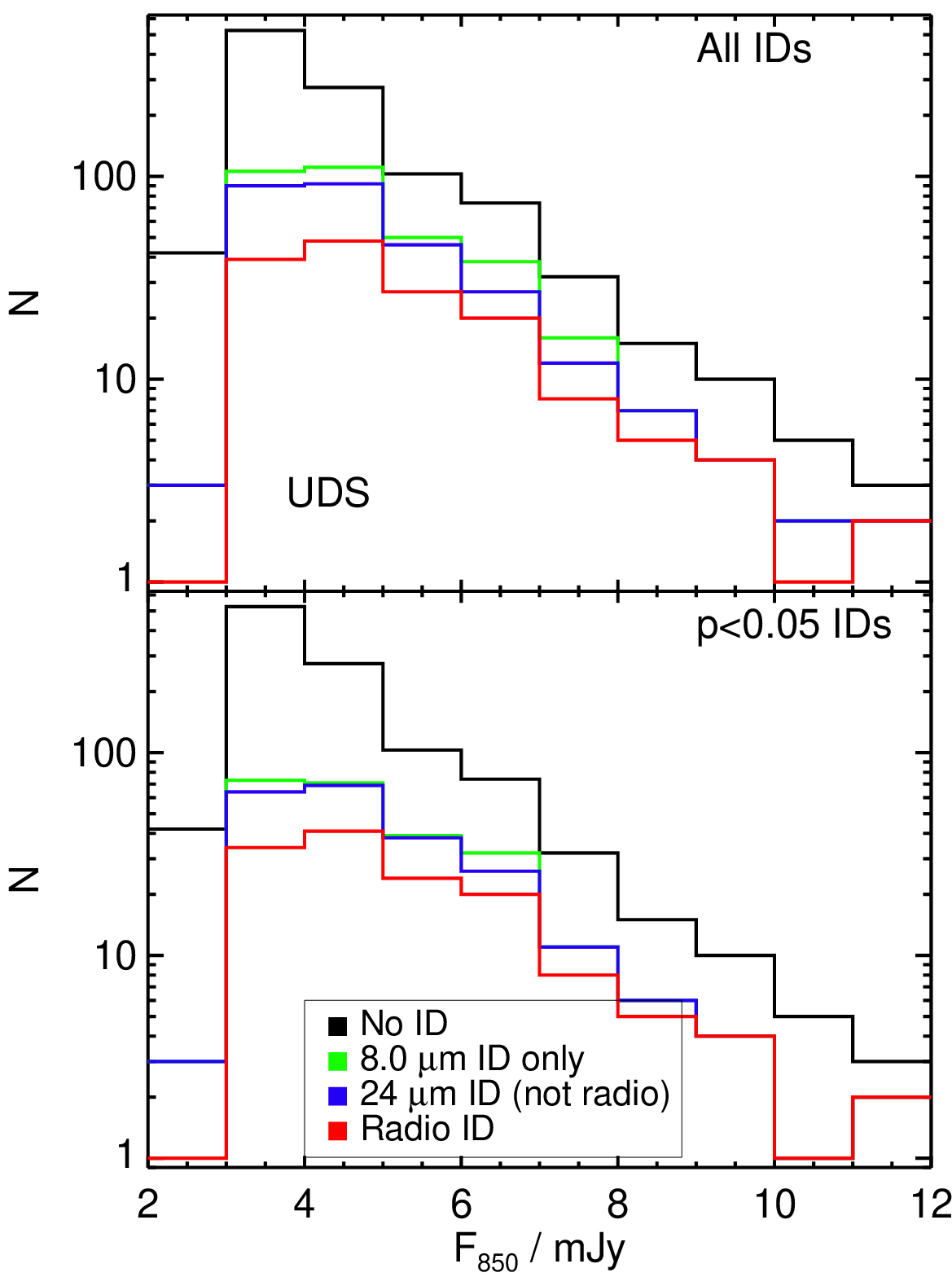}\\
\end{tabular}
 \caption{The number of IDs as a function of SCUBA-2 $850$-$\mu$m flux density
for the COSMOS and the 
UDS fields ({\it left} and {\it right}, respectively). 
 The {\it red histogram} shows the number of SCUBA-2 sources with radio IDs. 
 The space between the red histogram and the {\it blue histogram} shows the number of SCUBA-2 sources with $24$-$\mu$m IDs but no radio IDs.
 The space between the blue histogram and the {\it green histogram} shows the number of SCUBA-2 sources with only $8$-$\mu$m IDs.
 The space between the green histogram and the {\it black histogram} shows the number of SCUBA-2 sources with no IDs.
 The {\it upper panels} take into account all IDs, whereas the {\it lower panels} take into account only robust ($p\le0.05$) IDs.
The sharp decline in the number of sources in the UDS field below $3\,$mJy reflects the highly uniform 
depth of the SCUBA-2 map in this field. This map is also deeper than that of the COSMOS field. The histograms shown here contain all 1814 sources in the full $\ge 3.5\sigma$ SCUBA-2 sample.}
 \label{fig:Fsucc}
\end{figure*}

As in \citet{michalowski12}, we obtained the radio, $24\,\mu$m and $8\,\mu$m counterparts applying the method outlined in 
\citet{downes86}, \citet{dunlop89} and \citet{ivison07}. We applied a uniform search radius of 8\,arcsec, 
a conservatively high value in order to allow for astrometry shifts due to either pointing inaccuracies or source blending. 
This is an appropriate choice for the JCMT/SCUBA-2 $850\,\micron$ beam FWHM of $\simeq15\,$arcsec, as ALMA observations have 
revealed the brightest {\submm} sources up to approximately half the beam FWHM away from  the original JCMT/SCUBA-2 and 
APEX/LABOCA positions \citep{simpson15b,hodge13}. 

The statistical significance of each potential counterpart was assessed on the basis of the corrected Poisson probability $p$ that 
the chosen radio, $24\,\mu$m or $8\,\mu$m  candidate could have been selected by chance. IDs with a 
probability of chance association of $p\leq0.05$ are deemed to be `robust', whereas those with $0.05<p\leq0.1$ 
are labelled as `tentative'. If the $p$ values of multiple IDs for a given SCUBA-2 source 
satisfy these criteria, then all are retained, but the one with the 
lowest $p$ value is used for subsequent analysis.

IDs for the SCUBA-2 sources in the UDS field based on radio and optical counterparts 
(utilising an optical/near-IR colour selection) have previously been presented by \citet{chen16}. 
In this work, following our previous practice, we have complemented radio counterpart selection with searches for 
counterparts in the 24-$\mu$m and 8-$\mu$m {\it Spitzer} imaging. Nevertheless, the 
agreement between our IDs in the UDS field and those presented by \citet{chen16} is very good; restricting the SCUBA-2 
sample to the 716 $\ge4\sigma$ objects in the UDS field, only 90 of our robust  ($p\le0.05$) primary {\revone IDs (with the lowest $p$)} are not matched to 
those of \citet{chen16}, and 29 of these 90 are assigned $\mbox{{\it Class}}=2$ by \citet{chen16}, 
meaning that the optical data were inadequate for searching for IDs for these sources in the \citet{chen16} study.

{\revone All of our IDs for the $\ge 3.5\sigma$ 850\,$\mu$m sources in the COSMOS and UDS fields are presented in Tables~\ref{tab:IDCOSMOS} and \ref{tab:IDUDS} in the appendix, respectively.}

We summarise the outcome of the identification process in Table~\ref{tab:succ}, 
where we give the number of SCUBA-2 sources with IDs, and the nature 
of these IDs. We present the ID statistics split by ID wavelength and robustness, and 
also tabulate the results for three different significance cuts in the 850-$\mu$m source 
sample.
The number of IDs as a function of the SCUBA-2 $850\,\mu$m flux is plotted in Fig.~\ref{fig:Fsucc}
(shown here for the full $\ge 3.5\sigma$ SCUBA-2 sample). The ID rate is lower towards lower {\submm} fluxes. 
This is expected, both because of the increasing prevalence of false and/or flux-boosted sources at low 
significance, but also because
fainter {\smgs} have, on average, correspondingly lower radio and mid-IR fluxes
(as expected if the SED shape does not vary strongly from source to source;
see fig.~1 of \citealt{michalowski12} and of \citealt{ibar10}).

Unsurprisingly, the fraction of SCUBA-2 sources that lack IDs is also a function of 850-$\mu$m S/N.
As mentioned above, this is partly because the lower S/N sources are generally fainter, but also, as \citet{geach17} have shown from source injection and retrieval simulations, 
approximately 15--20 per cent of $\simeq3.5\sigma$ SCUBA-2 
sources located by the peak-finding method in these wide-area survey fields are either completely
erroneous or substantially flux-boosted. It is thus perhaps as expected that the 
unidentified fraction {\revone (with neither robust nor tentative IDs)} drops from $\simeq 35$ per cent at S/N $\ge$ 3.5 to $\simeq 20$ per cent at 
S/N $\ge$ 5.0 (where the percentage of false positive sources is expected to be $< 1$ per cent; \citealt{geach17}).
Despite this, we provide the IDs for all sources 
in the $3.5\sigma$ catalogue because, as Table~\ref{tab:succ} quantifies, the extended sample provides
a large number of additional robust identifications worthy of further study and follow-up. 
We therefore provide positions of all new IDs in the appendix. 

Nevertheless, it would clearly be wrong to infer that the real fraction of 
unidentified sources is as large as $\simeq 35$ per cent, when the evidence from the higher 
S/N cuts suggests the true figure is $\simeq 20-25$ per cent. Consequently, for the remainder 
of the analysis in this paper we consider only sources with S/N $\ge$ 4.0 (where 
the false positive SCUBA-2 source rate is expected to be $\simeq 5$ per cent; \citealt{geach17}).
At this S/N threshold, Table~\ref{tab:succ} shows we have robust IDs for $\simeq 60$ per cent of the 
1121 sources, with an additional $\simeq 15$ per cent having tentative IDs, and hence 
$\simeq 25$ per cent of sources remaining unidentified. About half of the robust IDs are provided
by the 1.4\,GHz radio imaging, and so extending the ID process to search for counterparts in the 
24-$\mu$m and 8-$\mu$m imaging 
has had a significant positive impact. We note that the ID statistics in the COSMOS and UDS fields are statistically consistent (due to the homogeneity of the 
SCUBA-2 data set, and the similar quality of the supporting data in the two survey fields).

In summary, we have completed the ID process and, for the $\simeq 1000$-source 
$\ge 4\sigma$ 850-$\mu$m sample, have identified $\simeq 75$ per cent of the sources. A key question, then, is why 
$\simeq 25$ per cent of the SCUBA-2 sources remain unidentified. 
There are several possible factors. First, some small remaining
subset of these sources may not be real. Secondly, as discussed further below, a few of these sources may in fact 
be blends of 2 or 3 significantly fainter sources, for which the optical/IR/radio counterparts lie below 
the flux-density limits of the supporting data; this is arguably not a serious problem since such sources 
should not really be retained in a bright flux-limited sample. Finally, some of the unidentified sources 
are likely to lie at higher redshifts where the resulting radio and mid-IR flux densities are too 
faint for their counterparts to be uncovered in the existing VLA and {\it Spitzer} imaging (which, unlike the 
submm imaging, does not benefit from a negative $k$-correction).
In the following sections we explore these issues further, first by revisiting the results of ALMA follow-up
of a subset of the SCUBA-2 sources, and then by exploiting the available long-wavelength (FIR--mm) data in the field
to attempt to constrain the redshifts of the unidentified SCUBA-2 sources.

\section{Comparison with ALMA follow-up}
\label{sec:alma}
\begin{figure}
  \includegraphics[width=0.45\textwidth]{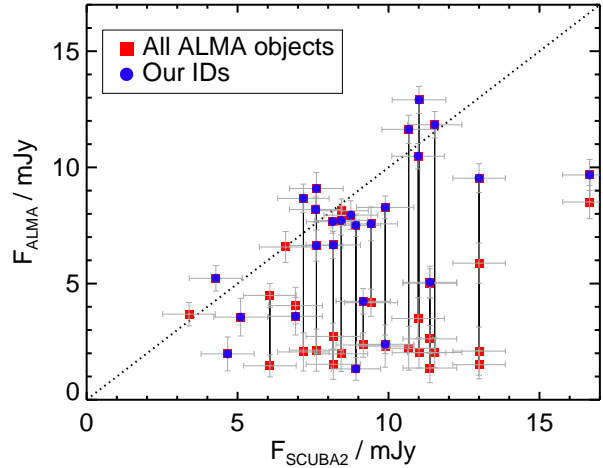}
  \caption{The ALMA flux densities for ALMA sources ({\it red and blue squares}) revealed through the 
follow-up of 29 SCUBA-2 sources in the UDS field \citep{simpson15b}, plotted 
against the SCUBA-2 single-dish flux density of each source as derived from the 
final SCUBA-2 CLS 850-$\mu$m imaging of the UDS field. The ALMA sources lying within 
{\revone the SCUBA-2 FWHM}
in the follow-up imaging are connected by {\it solid} vertical lines, and the 
{\it blue squares} indicate which of the ALMA sources was identified by our radio+mid-IR 
identification process as the location of the galaxy making the dominant contribution 
to the SCUBA-2 submm source. Although the brightest SCUBA-2 source divides into two ALMA 
subcomponents of comparable flux density, it can be seen that, in the vast majority of cases, 
the secondary ALMA component is a much fainter ($\simeq 1$--$2$\,mJy) object in the field. Moreover, the flux densities of the secondary components are not correlated with the brighter component flux densities, 
whereas the ALMA and SCUBA-2 flux densities of the brighter components are well correlated and 
frequently near equal (as indicated by the diagonal {\it dashed} line). For 25 of the 29 SCUBA-2 
sources, our radio+mid-IR identification process correctly locates the position of the dominant ALMA
component, yielding an estimated completeness of $\simeq 86$ per cent.}

 \label{fig:ALMAID}
\end{figure}

\subsection{Validation of galaxy identifications}
We can estimate the completeness and reliability of our identification procedure by considering
the subsample of 29 SCUBA-2 sources in our sample that has already been the subject of deep ALMA follow-up imaging \citep{simpson15b}. Although this subsample was originally selected to contain 
the brightest SCUBA-2 sources in the UDS field, the final deeper imaging from the S2CLS corrects for 
some of the more severe flux-boosting effects in the earlier map, with the consequence that this subset of sources actually contains objects with flux densities extending down to the flux-density limit of our sample
(and is thus more representative of the overall sample than originally anticipated).

In Fig.~\ref{fig:ALMAID} we plot the ALMA flux densities of 
{\revone all 52 ALMA galaxies versus the SCUBA-2 flux densities of the corresponding SCUBA-2 sources}
and  highlight in blue where, utilising the radio/mid-IR ID method adopted here, we have successfully located
the position of the galaxy counterpart as confirmed by ALMA. For many of the sources the ALMA imaging
has revealed more than one {\submm} component, and in Fig.~\ref{fig:ALMAID} we show this by connecting ALMA 
subcomponents with solid vertical lines. In the majority of cases it can be seen that the secondary
ALMA component is a much fainter ($\simeq 1$--$2$\,mJy) object in the field (i.e. lying within the 
{\revone SCUBA-2 FWHM}),
and that the flux densities 
of the secondary components are not correlated with the brighter component flux densities. For 
such faint submm galaxies we do not expect to be able to identify many galaxy counterparts given the
depth of the supporting imaging, but that is not a concern for this study that is focused 
on the study of sources with $S_{850} \ge 4$\,mJy. The key point is 
that our identification method has correctly identified the position of the {\it brighter} ALMA counterpart
for 25/29 of the sources, yielding a {\it completeness} of $\simeq 86$ per cent. 

\begin{figure}
\includegraphics[width=0.45\textwidth]{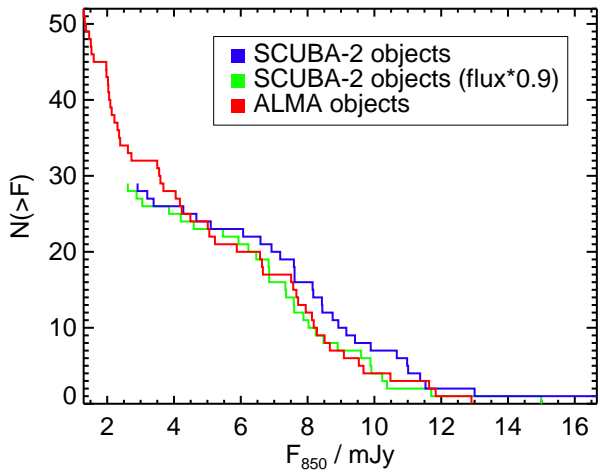}
 \caption{The cumulative distribution of single-dish JCMT/SCUBA-2 850-$\mu$m flux
densities ({\it blue}) and interferometric ALMA flux densities ({\it red}) resulting from the follow-up
imaging of 29 SCUBA-2 sources in the UDS field \citep{simpson15b}, as already illustrated in Fig.~\ref{fig:ALMAID}. 
Although the ALMA imaging reveals a population of fainter sources lying below the flux-density 
limit of the SCUBA-2 imaging, the bright end of the source counts is relatively little affected 
by whether one utilises the original SCUBA-2 flux densities, or those of the brighter ALMA subcomponents. Even without any correction, the flux distributions brightwards of $S_{850} \simeq 4$\,mJy are 
not significantly different (application of the Kolmogorov-Smirnov test yields a probability 
of only 50 per cent that  the ALMA and SCUBA-2 flux densities are not drawn from the same parent population),
but application of a modest correction, either subtracting $\simeq 1$\,mJy from all SCUBA-2 
flux densities or, as shown here, multiplying the SCUBA-2 flux densities by 0.9 ({\it green} distribution)
is sufficient to bring the SCUBA-2 and ALMA bright source counts into near perfect agreement.
}
 \label{fig:Fdistr}
\end{figure}

We can also use this control sample to estimate the {\it reliability} of our ID method (i.e. the fraction of IDs confirmed by ALMA). 
{\revone We have identified 25 robust IDs (21 primary) for the 29 SCUBA-2 sources with ALMA follow-up. Of these, 23 (20) are confirmed by ALMA, while two are not. This yields a {\it reliability} of $\simeq92$ per cent that our primary galaxy identifications correspond to submm sources. We have also identified six tentative ID (three primary), out of which two were confirmed by ALMA.}

It might be argued that these estimates of completeness and reliability could be optimistic, because the 
ALMA control sample utilised here remains biased towards higher {\submm} flux densities than the full
UDS and COSMOS samples. However, Fig.~\ref{fig:ALMAID} shows that we correctly identified three of the four faintest 
sources in the control sample, and we re-iterate that we are not concerned with identifying 
sources (either SCUBA-2 sources, or ALMA subcomponents) significantly fainter than $S_{850} \simeq
4$\,mJy. Moreover, the high success rate of the radio+mid-IR identification approach has 
already been confirmed by the ALMA follow-up of the LABOCA sources in the LESS survey, as 
described by \citet{hodge13}. Despite the significantly larger beam delivered by the LABOCA imaging as 
compared to the SCUBA-2 imaging (approaching a factor of 2 in beam area, with 
thus significantly increased likelihood of source multiplicity and blending), 
45 out of 57 of the robust  IDs for LABOCA sources proposed by \citet{biggs11} were
confirmed by the ALMA imaging \citep{hodge13}, yielding a reliability of $\simeq 80$ per cent, and 
the correct position of the brightest ALMA component was correctly predicted by the radio ID 
for 52 out of 69 LABOCA sources, yielding a completeness of $\simeq 75$ per cent.
{\revone This is higher than the completeness quoted by \citet{hodge13}, but they included all ALMA sources, not just the brightest ones for each LABOCA source. Our approach gives the fraction of single-dish sources for which the main component was correctly identified.}

\subsection{Multiplicity and number counts}

First with the IRAM PdB and the SMA, and more recently with ALMA, it has now become possible to address the issue of the extent to which the {\smgs} detected by single-dish surveys consist of blends of fainter {\submm} galaxies lying within the single-dish primary beam \citep{wang11,hodge13,karim13,simpson15b}. 
{\revone Most of these studies reported a very high ($>50$\%) multiplicity rate, but this was based on including even the faintest {\submm} companions in the statistics, and in several cases the single-dish beamsize was also significantly larger than delivered by the JCMT at 850\,\micron. In what follows we revise these numbers by treating as multiple only the cases when the secondary companion is sufficiently bright to potentially affect the identifications if only single-dish observations were available.}
The impact of real physical associations, or simply the blending of projected sources (i.e. at 
very different redshifts) on single-dish flux densities and derived number counts is obviously 
a function of the size (i.e. FWHM) of the single dish primary beam, and hence is more serious for 
surveys conducted at longer wavelengths, or with smaller telescopes.

In the LABOCA  Extended Chandra Deep Field South {\submm} survey \citep[LESS;][]{weiss09b}, 20 out of 69 LABOCA {\submm} sources ($\simeq 30$ per cent) were revealed by ALMA follow-up imaging
to comprise multiple ALMA sources with a flux-density ratio $<3$, leading to suggestions
that source multiplicity might be a serious problem for previous single-dish
{\submm} surveys \citep{hodge13,karim13} (a flux-density ratio threshold of 3 is usually adopted as a minor/major merger threshold, e.g. \citealt{lambas12}; and also provides a reasonable threshold for considering which sources have had their single-dish flux-densities and positions seriously affected by source multiplicity/blending). A slightly smaller fraction (16 of these 69, $\equiv 23$ per cent) 
of these sources were also found to have multiple radio IDs \citep{biggs11}. 

However, the beam area of APEX/LABOCA is nearly twice as large as that of JCMT/SCUBA-2, so the 
impact of source multiplicity on the SCUBA-2 results is expected to be significantly smaller.
This is confirmed by the ALMA follow-up of the SCUBA-2 sources by \citet{simpson15b} as 
already presented in Fig.~\ref{fig:ALMAID}. Here only 6 out 30 ($\simeq20$ per cent) of the SCUBA-2 sources have been 
found to consist of multiple ALMA sources with a flux-density ratio $<3$, and arguably this is an overestimate for the full SCUBA-2 source sample, given that the sample studied by \citet{simpson15b}
is biased towards brighter sources where blending is likely to be a more serious issue {\revone (due to the steep slope at the bright end of the {\submm} luminosity function)}. Indeed, as is evident
from Fig.~\ref{fig:ALMAID}, while the brightest SCUBA-2 source is clearly revealed to be a blend of two ALMA 
components with comparable flux densities, the majority of SCUBA-2 source flux densities are well 
matched by the flux densities of the brighter ALMA components, and it is clear that in most cases 
the secondary ALMA component is either too faint, or too well-separated from the brighter component
to significantly contaminate/bias the SCUBA-2 derived flux density.

We can also explore the issue of multiplicity from a radio perspective, by considering the prevalence 
of multiple radio IDs within the SCUBA-2 sample.
If a {\submm} source is composed of two or more sources with similar luminosities at similar redshifts, then if the primary component is detected in the radio with high signal-to-noise ratio, then the secondary component should also be detected. However, in the COSMOS (UDS) field, out of 181 (332) SCUBA-2 sources with radio IDs, only 14 (26) have multiple radio IDs, i.e.~$\simeq8$ per cent ($8$ per cent).
{\revone For  14 (18) of them the secondary ID is also robust.}
The corresponding numbers for multiple $24$-$\mu$m 
IDs are 7 per cent, 
{\revone or  27/401, 8 with robust secondary IDs} 
(6 per cent {\revone or  33/587, 5 with robust secondary IDs}). 
Finally,  9 per cent 
{\revone or 30/326, 5 with robust secondary IDs} 
(10 per cent {\revone or  45/452, 6 with robust secondary IDs}) of  $8$-$\mu$m IDs are multiple. 
However, the true multiplicity rate is likely higher, because unidentified sources
could also represent blends of fainter {\submm} sources. Hence, to better assess the ID multiplicity rate, we confined our attention to a subsample with high radio ID completeness. 
Among the twenty $\ge10\sigma$ SCUBA-2 sources in the UDS field 17 (85 per cent) have radio IDs 
and only 2/17 of these (i.e. 12 per cent) are multiple. An upper limit on multiplicity can be derived 
by assuming that all sources lacking a radio are multiple, yielding $(2+3)/20$ (i.e. 25 per cent).  
 
We conclude that, within our SCUBA-2 sample, only $\simeq 15$--$20$ per cent of sources are potentially significantly affected by multiplicity and blending. Moreover, as we show in Fig.~\ref{fig:Fdistr}, the impact 
of any multiplicity/blending on the bright-end of the 850-$\mu$m source counts as derived 
from SCUBA and SCUBA-2 surveys with the JCMT is very modest. This shows that, even without any correction, the SCUBA-2 and ALMA flux-density distributions brightwards of $S_{850} \simeq 4$\,mJy are 
not significantly different (application of the Kolmogorov-Smirnov test yields a probability 
of only 50 per cent that  the ALMA and SCUBA-2 flux densities are not drawn from the same parent population),
and that application of a modest correction, either subtracting $\simeq 1$\,mJy from all SCUBA-2 
flux densities or, as shown in Fig.~\ref{fig:Fdistr}, multiplying the SCUBA-2 flux densities by 0.9,
is sufficient to bring the SCUBA-2 and ALMA bright source counts into near perfect agreement.
Our findings on the small impact of multiplicity on number counts are in agreement with those of \citet{chen13b}, in which they found only $\sim15$\% of their SMA-targeted SCUBA-2 {\submm} sources are multiples, and therefore their SCUBA-2 counts are not significantly affected by multiplicity either.
Previous claims that submm number counts have been severely biased by source blending 
appear to have been exaggerated, and in any case have generally been based on samples derived from
imaging surveys with much larger beam sizes than are provided by the JCMT at 850 $\mu$m
\citep{karim13}.

To summarise, given the success of our ID procedure in locating the positions of the brightest 
ALMA components, the relatively low prevalence of significant ALMA subcomponents or secondary radio 
IDs, and the modest impact of source multiplicity on the bright end of the 850-$\mu$m source counts, 
it is clear that source multiplicity and blending is not a serious issue for the study of bright
850-$\mu$m sources selected at the angular resolution provided by the JCMT.

\section{Long-wavelength photometry}
\label{sec:phot}

We now return to the issue of completing the redshift content of the SCUBA-2 sample, and determining the physical properties of the sources.
Because $\simeq 25$ per cent of even the $\ge4\sigma$ SCUBA-2 sources remain unidentified at optical/near-IR/mid-IR/radio wavelengths, and
because some of the optical identifications may be wrong (either because they are not statistically robust, or because they
are intervening lenses) it is crucial to utilise the available far-infrared and mm imaging available in the field
to enable at least crude constraints on redshift to be established (by fitting to the anticipated rest-frame far-infrared SED of the dust emission).
This information is also important, even for the identified sources, for estimating the dust-enshrouded SFR of each object.

We therefore used the {\it Herschel}\footnote{{\it Herschel} is an ESA space observatory with science instruments provided by
European-led Principal Investigator consortia and with important participation from NASA.} \citep{herschel}
Multi-tiered Extragalactic Survey \citep[HerMES;][]{oliver12,levenson10,viero13} and the PACS Evolutionary Probe \citep[PEP;][]{lutz11}
data  obtained with  the Spectral and Photometric Imaging Receiver \citep[SPIRE;][]{spire} and  the Photodetector Array Camera and Spectrometer  \citep[PACS;][]{pacs}, covering the entire COSMOS and UDS fields. We used maps at $100$, $160$, $250$, $350$ and $500\,\mu$m with beam sizes of $7.4$, $11.3$, $18.2$, $24.9$ and $36.3$\,arcsec. 
The maps are available through the {\it Herschel} Database in Marseille (HeDaM)\footnote{\url{hedam.lam.fr}} and the
PEP website\footnote{\url{www.mpe.mpg.de/ir/Research/PEP/DR1}}.

In addition, in order to constrain the long-wavelength side of the SEDs of SCUBA-2 sources, we used the $1.1$\,mm
AzTEC imaging data available in both survey fields. This imaging unfortunately does not cover all of the area surveyed with
SCUBA-2, and is less deep than is desirable, but nevertheless is provides detections for some of our $850\,\micron$-selected galaxies,
and useful upper limits for a significant fraction of the remainder.
For the COSMOS field we used the JCMT and ASTE AzTEC \citep{aztec}
maps and catalogues from \citet{scott08}, and \citet{aretxaga11}, covering $0.15$ and $0.72$\,deg$^2$ down to an rms of $1.3$ and $1.26$\,mJy\,beam$^{-1}$, respectively. 
For the UDS field we used the JCMT and ASTE AzTEC data from \citet{austermann10} and Kohno (private communication). These  cover
$0.7$ and $0.27$\,deg$^2$ to an rms depth of $1.0$--$1.7$ and $0.5$\,mJy\,beam$^{-1}$, respectively. 

We obtained the {\it Herschel} fluxes of each SCUBA-2 source in the following way.
We extracted 120-arcsec wide stamps from all five {\it Herschel} maps around the position of each SCUBA-2 source.
Then we processed the PACS ($100$ and $160\,\mu$m) maps  by simultaneously fitting Gaussians with
the FWHM of the respective maps, centred at the positions of all radio and $24$-$\micron$ sources located
within these cut-outs, and at the positions of the SCUBA-2 IDs. 
Then, to deconvolve the SPIRE ($250$, $350$ and $500\,\mu$m)
maps in a similar way, we used the positions of the $24$-$\mu$m sources detected with PACS ($\ge3\sigma$),
the positions of all radio sources, and the SCUBA-2 ID positions (or the {\submm} positions if no
radio or mid-IR ID had been secured). 
The errors were computed from the covariance matrix of the fit, in which the free parameters are simply the heights of the Gaussian
beams fitted at each input position. Then the confusion noise of 5.8, 6.3 and 6.8\,mJy\,beam$^{-1}$ at 250, 350 and 500\,\micron,
respectively \citep{nguyen10} was added in quadrature. The fitting was performed using the IDL
{\sc Mpfit}\footnote{\url{purl.com/net/mpfit}} package  \citep{mpfit}.

To incorporate the information from the AzTEC imaging, we
matched the SCUBA-2 and 1.1\,mm catalogues within 12\,arcsec
(the approximate sum in quadrature of the positional uncertainties
of SCUBA-2 and AzTEC sources), which resulted in 72 matches in the
COSMOS field, and 118 matches in the UDS field. Then we estimated the
1.1\,mm fluxes for the non-matched SCUBA-2 sources in the same way as
for the {\it Herschel} fluxes. This was possible for an
additional 211 SCUBA-2 sources in the COSMOS field and 250 SCUBA-2 sources
in the UDS field.

The derived long-wavelength fluxes are presented in
Tables~\ref{tab:IDFlongCOSMOS} and \ref{tab:IDFlongUDS} in the appendix.

Because the 1.1-mm information proves to be crucial for setting
meaningful upper bounds on the `long-wavelength' redshift estimates
(particularly for SCUBA-2 sources with weak, or non-existent {\it Herschel} detections),
we have restricted the remainder of the analysis presented in this paper to the subset
of 651 (out of 1121) $\ge 4\sigma$ SCUBA-2 sources for which the AzTEC 1.1-mm
coverage is available {\revtwo (283 in the COSMOS field and 368 in the UDS field).}


\section[]{Redshifts and number density}
\label{sec:z}


\begin{figure}
\includegraphics[width=0.45\textwidth]{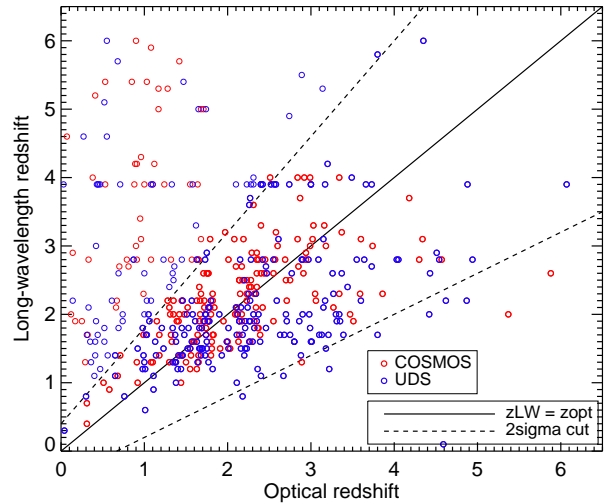}
 \caption{Long-wavelength photometric redshift as a function of optical/near-IR photometric redshift  (Section~\ref{sec:z}) for the $\geq4\sigma$ SCUBA-2 sources in the COSMOS ({\it red}) and UDS ({\it blue}) fields that have $1.1$\,mm coverage and optical/near-IR galaxy counterparts.
The {\it solid line} represent $z_{\rm LW}=z_{\rm opt}$, whereas the {\it dashed lines} show the $2\sigma$ cut from the Gaussian fit presented in Fig.~\ref{fig:dz}. Thinner symbols (above the  upper dashed line) represent objects for which the long-wavelength redshifts were adopted (see Section~\ref{sec:z}). The concentration of points at $z_{\rm LW}=3.9$ is due to {\it Herschel}-undetected objects for which the minimum $\chi^2$ yielded by the long-wavelength fitting
is almost flat above some lowest permitted 
value, and the formal best-fitting solution is at that lowest allowed redshift.
}
 \label{fig:zz}
\end{figure}

\begin{figure}
\includegraphics[width=0.45\textwidth]{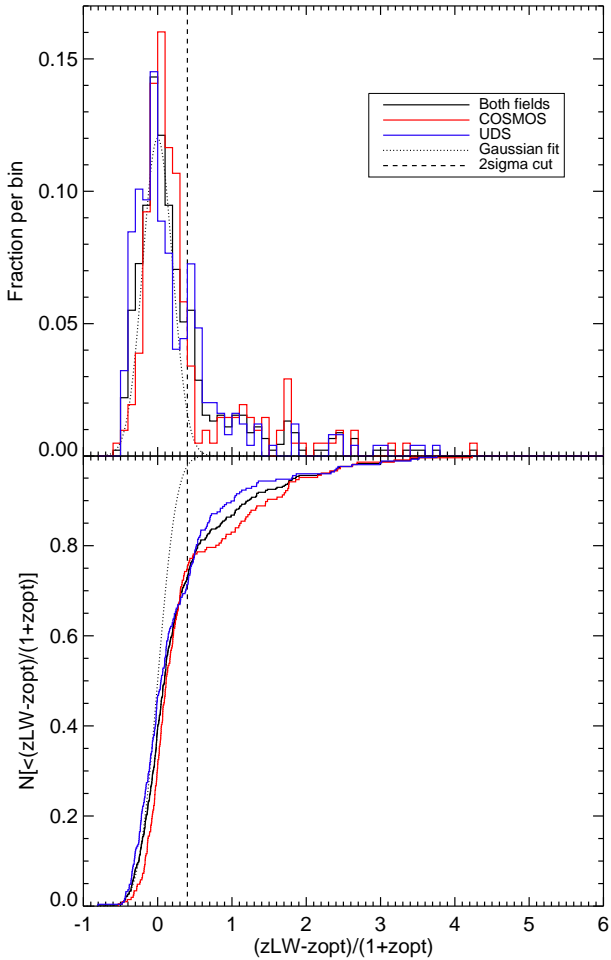}
 \caption{The distribution ({\it upper panel}) and cumulative distribution ({\it lower panel}) of the difference between the long-wavelength and optical/near-IR photometric redshifts (Section~\ref{sec:z}) for the  $\geq4\sigma$ SCUBA-2 sources that have $1.1$\,mm coverage and optical/near-IR galaxy counterparts. The {\it solid curves} are colour-coded depending on the field. The {\it dotted line} is a Gaussian fit to the negative side of distribution (with $\sigma=0.23$), whereas the {\it dashed line} is the $2\sigma$ cut of this Gaussian, above which the optical redshifts are deemed incorrect due to poorly determined redshifts, incorrect identifications, or because the optical counterpart is a likely a foreground  galaxy lens.}
 \label{fig:dz}
\end{figure}

\begin{figure}
\includegraphics[width=0.45\textwidth]{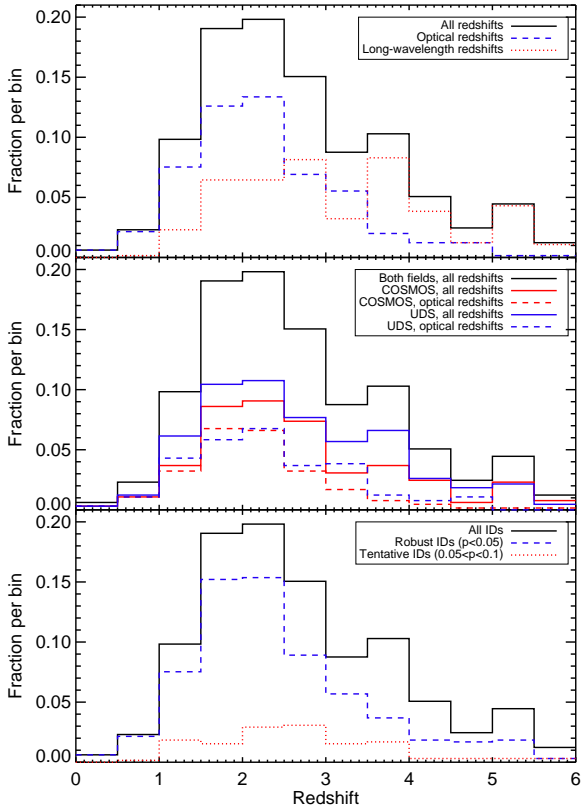}
 \caption{{\it Top}: the redshift distribution of the $\geq4\sigma$ SCUBA-2 sources that have $1.1$\,mm coverage showing all sources ({\it black solid line}), those with optical/near-IR redshifts ({\it blue dashed line}), and those with long-wavelength redshifts only ({\it red dotted line}). 
 {\it Middle}: the redshift distribution divided by field. The line type is the same as in the top panel: {\it solid lines} denote all redshifts, and {\it dashed lines} denote optical/near-IR redshifts. 
The lines are colour-coded by the field: {\it black}: both fields, {\it red}: COSMOS, {\it blue}: UDS.
 {\it Bottom}: the redshift distribution divided by the quality of IDs. The {\it black solid line} is the same as above, whereas the {\it blue dashed line} denotes robust IDs ($p\leq0.05$), and the {\it red dotted line} denotes tentative IDs ($0.05<p\leq0.1$).}
 \label{fig:z}
\end{figure}

\begin{figure}
\includegraphics[width=0.45\textwidth]{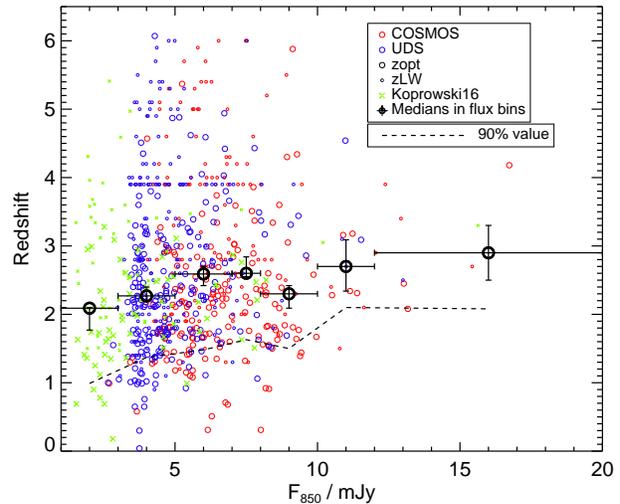}
 \caption{Redshift as a function of  $850\,\micron$ flux density for the SCUBA-2 sources in the COSMOS ({\it red}) and UDS ({\it blue}) fields, and in the deeper data in the COSMOS field ({\it green crosses}) presented in \citet{koprowski16}. Again we plot only the $\geq4\sigma$ SCUBA-2 sources that have $1.1$\,mm coverage. Larger symbols correspond to sources with optical/near-IR redshifts, whereas smaller symbols indicate those with only long-wavelength redshifts. The {\it black circles with error bars} correspond to medians in flux bins indicated by horizontal error bars. The {\it dashed line} shows the redshift value above which 90 per cent of objects in a given flux-density bin are located.  
The correlation of flux and redshift  is significant, as the Spearman's rank correlation coefficient is $0.19$ with a very small probability ($\sim3\times10^{-7}$) that the null hypothesis (no correlation) is correct.
Sources at $z=6$ have only {\it Herschel} upper limits, so while the formal best solution is at the maximum allowed redshift, the error bars are large, and extend to much lower redshifts. On the other hand, the concentration of points at $z_{\rm LW}=3.9$ is due to {\it Herschel}-undetected objects for which the redshift-dependence of minimum $\chi^2$ is nearly flat above some minimum permitted value, and the formal best solution is at that lowest redshift.}
 \label{fig:850z}
\end{figure}

\begin{table*}
\caption{Median properties of $\geq4\sigma$ SCUBA2 sources with 1.1\,mm coverage.}
\label{tab:sfr}
\begin{center}
\begin{tabular}{lccccccc}
\hline\hline
Field & $z$ & $z_{\rm opt}$ & SFR & $\mbox{SFR}_{z_{\rm opt}}$ & $\log(\mstar/\msun)$ & sSFR & $f_{\rm old}$ \\
    &     &     & (\msunyr) & (\msunyr) &  & (\pergyr) \\
(1) & (2) & (3) & (4) & (5) & (6) & (7) & (8) \\
\hline
COSMOS & $2.40^{+0.11}_{-0.05}$ & $2.11^{+0.04}_{-0.14}$ & $324^{+8}_{-10}$ & $301^{+5}_{-14}$ & $11.11^{+0.05}_{-0.04}$ & $2.12^{+0.13}_{-0.17}$ & $0.75^{+0.08}_{-0.03}$ \\
UDS & $2.42^{+0.17}_{-0.06}$ & $2.24^{+0.03}_{-0.04}$ & $261^{+6}_{-6}$ & $258^{+6}_{-6}$ & $11.13^{+0.02}_{-0.01}$ & $1.97^{+0.14}_{-0.11}$ & $0.93^{+0.01}_{-0.01}$ \\
Both &$2.40^{+0.10}_{-0.04}$ & $2.17^{+0.04}_{-0.04}$ & $287^{+7}_{-6}$ & $269^{+13}_{-7}$ & $11.12^{+0.02}_{-0.02}$ & $2.02^{+0.10}_{-0.08}$ & $0.91^{+0.01}_{-0.02}$ \\
\hline
\end{tabular}
\end{center}
{(1) Field name; (2) Median redshift including all sources; (3) Median optical photometric redshift; (4) Median SFR including all objects (using long-wavelength redshifts if optical redshifts are not available); (5) Median SFR including only objects with optical photometric redshifts; (6)  Median stellar mass; (7) Median specific SFR. (8) Fraction of stellar mass contributed by the old stellar component (see Section~\ref{sec:z}). For all properties but redshift only objects at $z>1$ were taken into account.}
\end{table*}

We used the optical, near-IR and IRAC data (presented in Tables~\ref{tab:IDoptCOSMOS} and \ref{tab:IDoptUDS} in the appendix) to fit the SEDs of all IDs and to derive their photometric redshifts and physical properties using the method of  \citet{cirasuolo07,cirasuolo10}.
This uses a modification of  the {\sc HyperZ} package \citep{bolzonella00} 
with the stellar population models of \citet{bruzualcharlot03} and a \citet{chabrier03} initial 
mass function (IMF) with a mass range $0.1$--$100\,$M$_\odot$. 
A double-burst star formation history was assumed, but this choice has little impact on derived redshifts \citep[as opposed to derived stellar masses, which are well reproduced by the two-component star formation history for {\smgs};][]{michalowski12mass,michalowski14}. 
The metallicity was fixed at the solar value and 
reddening was calculated following the \citet{calzetti00} law within the range $0 \le A_V	\le 6$ \citep[see][]{dunlop07}. 
The age of the young stellar component was varied between  $50$\,Myr and $1.5$\,Gyr, and the old component was allowed to contribute $0$--$100$\ per cent of the near-IR emission while its age was varied over the range $1$--$6$\,Gyr.
The HI absorption along the line of sight was included according to the prescription of \citet{madau95}.
The accuracy of the photometric catalogue of \citet{cirasuolo10} is excellent,  with a mean $|z_{\rm phot}-z_{\rm spec}|/(1+z_{\rm spec})=0.008\pm0.034$. 

We also estimated `long-wavelength' redshifts, as in \citet{koprowski14,koprowski16}, fitting the average {\smg} template \citep[from][]{michalowski10smg} to the far-IR and (sub){\mm} photometry ({\it Herschel} PACS, SPIRE, SCUBA-2 and AzTEC data). 
{\revone Non-detections were treated in the same way as detections in the fitting, using the flux and error measured at a given position. Hence, the case of {\it Herschel} non-detections resulted in ruling out low-$z$ solution and flat $\chi^2$ distributions  at higher redshifts.}
Long-wavelength redshifts were especially useful for sources with no optical counterparts (or no IDs at all).
This redshift determination is obviously not as accurate as the optical photometric method, but provides an
important estimate of the $\Delta z \simeq 0.5$-wide redshift bin within which a given source resides. For sources with optical/near-IR redshifts the 
median $|z_{\rm LW}-z_{\rm opt}|/(1+z_{\rm opt})$ for the COSMOS field is  $\simeq 0.16\pm0.03$, 
while for the UDS field it is $\simeq0.011\pm0.016$. This is similar to the accuracy reported in \citet{aretxaga05,aretxaga07}.

{\revone For both redshift estimates the errors were calculated by the determination of the redshift range over which $\chi^2$ increases by 1 from the minimum value while allowing all other parameters to vary.}

The resulting redshifts are given in Tables \ref{tab:IDzCOSMOS} and \ref{tab:IDzUDS} in the appendix.
For sources with multiple IDs, the ID with the smallest $p$-value was used. 
The fraction of SCUBA-2 sources with optical/near-IR redshifts is summarised in Table~\ref{tab:succ} (column 16). 
We obtained optical/near-IR photometric redshift estimates for $\simeq60$ per cent of the SCUBA-2 sources located inside the deep optical/near-IR imaging maps. The remaining $\simeq40$ per cent either do not have IDs at all, or no optical source was matched to the radio/mid-IR IDs.

{\revone For 50 IDs in the COSMOS field and 20 in the UDS spectroscopic redshifts (Section~\ref{sec:dataopt}) were available and used instead of optical photometric redshifts.}

As in \citet{koprowski16},  we attempted to filter the optical/near-IR redshifts, replacing these redshift
estimates with the long-wavelength photometric redshift values when the two values are formally inconsistent. In practice, where the two values differ dramatically, it is in the sense that the optical/near-IR photometric redshift estimate is too low, either because the optical counterpart has been assigned
in error, or because the identified optical galaxy is in fact lensing a more distant submm source  
\citep[as in][]{negrello10}. In Fig.~\ref{fig:zz} we show the long-wavelength redshift as a function of optical redshift, and in Fig.~\ref{fig:dz} we show the distribution of the difference between the long-wavelength and optical redshifts, $(z_{\rm LW}-z_{\rm opt})/(1+z_{\rm opt})$. 

We fitted a Gaussian to the negative side of the distribution obtaining a width of $\sigma=0.23$. Then we discarded optical/near-IR photometric 
redshifts (and the corresponding IDs) for sources with long-wavelength 
redshifts that are $2\sigma$ higher (above the dashed lines in Figs~\ref{fig:zz} and \ref{fig:dz}), and thereafter retain only the long-wavelength redshift estimates for these sources.
{\revone This happened for 23 robust and 14 tentative primary IDs in the COSMOS field and 42 robust and 11 tentative primary IDs in the UDS field.}
{\revtwo  Out of 651 $\geq4\sigma$ sources with 1.1\,mm coverage 349 have optical counterparts retained in the analysis because of the consistency with the long-wavelength redshift  (160 in the COSMOS field and 189 in the UDS field).}

{\revtwo
The substantial scatter in the $z_{\rm LW}$ versus $z_{\rm opt}$ plot 
(Fig.~\ref{fig:zz}) can be fully explained by photometry measurement errors. The median contribution of the data points to the $\chi^2$ with respect to the  $z_{\rm LW}=z_{\rm opt}$ line is $\sim1.5$, so this model explains the data reasonably well. This justifies our choice of a single template in deriving  $z_{\rm LW}$, as the data do not require a more complex model.
}

The resulting final redshift distribution of the SCUBA-2 sources is shown in Fig.~\ref{fig:z}, sub-divided by the type of redshift calculation (all, optical/near-IR, long-wavelength), by the survey field, and by the 
quality of the ID. 
The median redshift for the full $\ge 4\sigma$ SCUBA-2 sample with 1.1-mm coverage is 
$z = 2.40^{+0.10}_{-0.04}$ for all sources, or $z = 2.17\pm 0.04$ for the subset of sources 
with retained optical/near-IR redshifts (see Table~\ref{tab:sfr}), 
 consistent with previous studies of smaller samples of {\smgs} \citep{chapman05,chapin09,wardlow11,michalowski12,yun12,simpson14,chen16,koprowski16}. 

While the median redshifts are consistent with previous studies, our large sample size, 
and the use of long-wavelength photometric redshifts to complete the redshift
content of the sample, has enabled us to more clearly reveal/define the extent of the 
high-redshift tail of the submm galaxy population. Obviously, sources with no optical/near-IR redshifts (red dotted histogram on the top panel of Fig.~\ref{fig:z}) have, on average, higher redshifts than the remaining sample. Taking into account both optical/near-IR
and long-wavelength redshifts, as much as 393 out of  1691 (23 per cent) 
SCUBA-2 sources are at $z\geq4$.  However, only 39 sources (10 in COSMOS and 29 in UDS field) have {\it optical/near-IR} $z\geq4$. 
Similarly, out of  
{\revone 651 $\geq4\sigma$ SCUBA-2 sources with 1.1\,mm coverage, 93 (14 per cent) are at $z\geq4$ and 19 have optical/near-IR $z\geq4$  (6 in COSMOS and 13 in the UDS field).}

The middle panel of Fig.~\ref{fig:z} shows that the redshift distributions in the COSMOS and UDS fields separately (both using all redshifts and only optical redshifts) are qualitatively similar, displaying a peak at $z\simeq2$. This means that with $\simeq1\,\mbox{deg}^2$ fields we start to overcome the cosmic variance, which makes number counts \citep{scott10,scott12} and redshift distributions \citep{michalowski12} derived using smaller fields significantly different from each other. 
Application of the Kolmogorov-Smirnov test results in probability of $\simeq0.7$ per cent
that the COSMOS and UDS samples are drawn from the same parent population, but this is a $\lesssim3\sigma$ discrepancy.

The redshift distribution of tentative IDs  ($0.05<p\leq0.1$, red dotted histogram on the bottom panel of Fig.~\ref{fig:z}) is not shifted towards lower redshifts with respect to robust IDs ($p\leq0.05$, blue dashed line), as would be expected if tentative IDs were significantly contaminated by unrelated galaxies (because lower-redshift galaxies dominate optical catalogues). In any case, the fraction of tentative IDs is only $\simeq15$ per cent (Table~\ref{tab:succ}; {\revone both before and after long-wavelength redshift filtering}), so they do not significantly affect our conclusions.

It has been suggested in the past that {\smgs} with higher fluxes are located at preferentially at higher redshifts \citep{ivison02,pope05,michalowski12,koprowski14}, and with our large sample we are
able to further investigate this issue.  Fig.~\ref{fig:850z} shows {\submm} flux as a function of redshift for the SCUBA-2 sources presented here and in a deeper SCUBA-2 image in the COSMOS field \citep{koprowski16}.  It is evident that the bottom-right corner of this plot (high flux density, low redshift) is empty, and this is not due to selection effects, as such sources should be easy to detect at all wavelengths, and redshifts easy to measure. The scatter in this figure is large but a weak overall trend can be discerned. 
The Spearman rank correlation coefficient is $0.19$ with a very small probability ($\sim3\times10^{-7}$) of the null hypothesis (no correlation) being acceptable. However, there is no real evidence for a 
deficit of lower luminosity objects at high-redshift, and so this statistically significant correlation is 
driven by the absence of submm bright low-redshift objects; very luminous submm galaxies are only found 
in our survey at $z > 2$.

\begin{figure}
\includegraphics[width=0.45\textwidth]{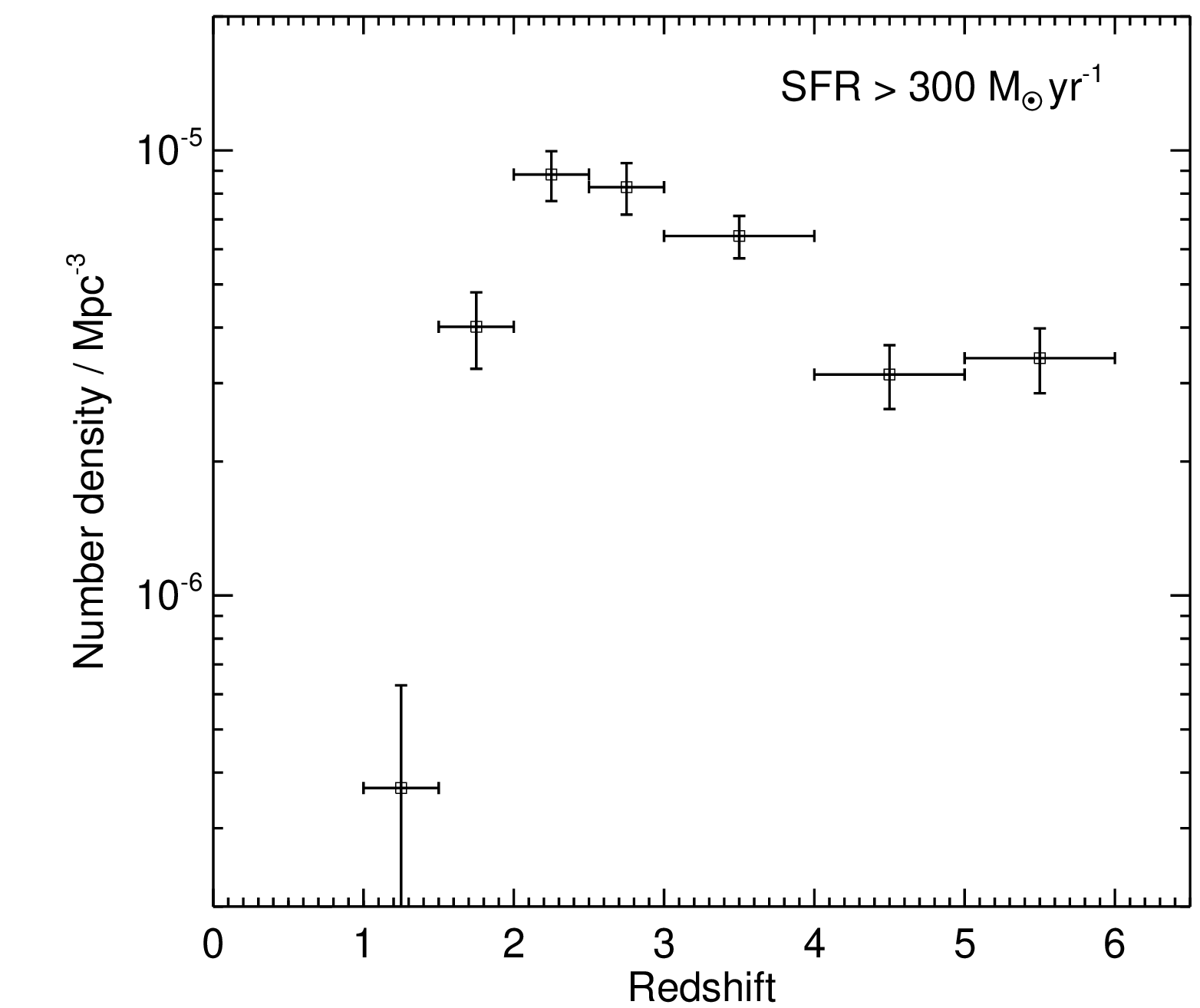}
 \caption{The comoving number density as a function of redshift of {\smgs} with $\mbox{SFR}\geq300\,\msunyr$ (our survey is sensitive to such objects at all redshifts, see Fig.~\ref{fig:sfr}). 
}
 \label{fig:nden}
\end{figure}

In Fig.~\ref{fig:nden} we utilise the redshift content of our SCUBA-2 sample to plot the 
comoving number density of {\smgs} with $\mbox{SFR}\geq300\,\msunyr$ as function of redshift. {\revone The values are shown in Table~\ref{tab:fracsmg}.} Our survey is sensitive to such objects at all redshifts (see next section, and Fig.~\ref{fig:sfr}), so this figure shows an unbiased and complete estimate of the cosmological evolution of the number density of the most
luminous star forming galaxies in the Universe. It can be seen that, although such objects are largely 
confined to $z > 2$, their number density declines significantly beyond $z \simeq 3.5$. Nevertheless,  
they still appear to persist at number densities significantly in excess of $10^{-6}\,\mbox{Mpc}^{-3}$ at $ z \simeq 5$.

\section{Star formation rates and stellar masses}
\label{sec:sfr}

\begin{figure*}
\includegraphics[width=0.8\textwidth]{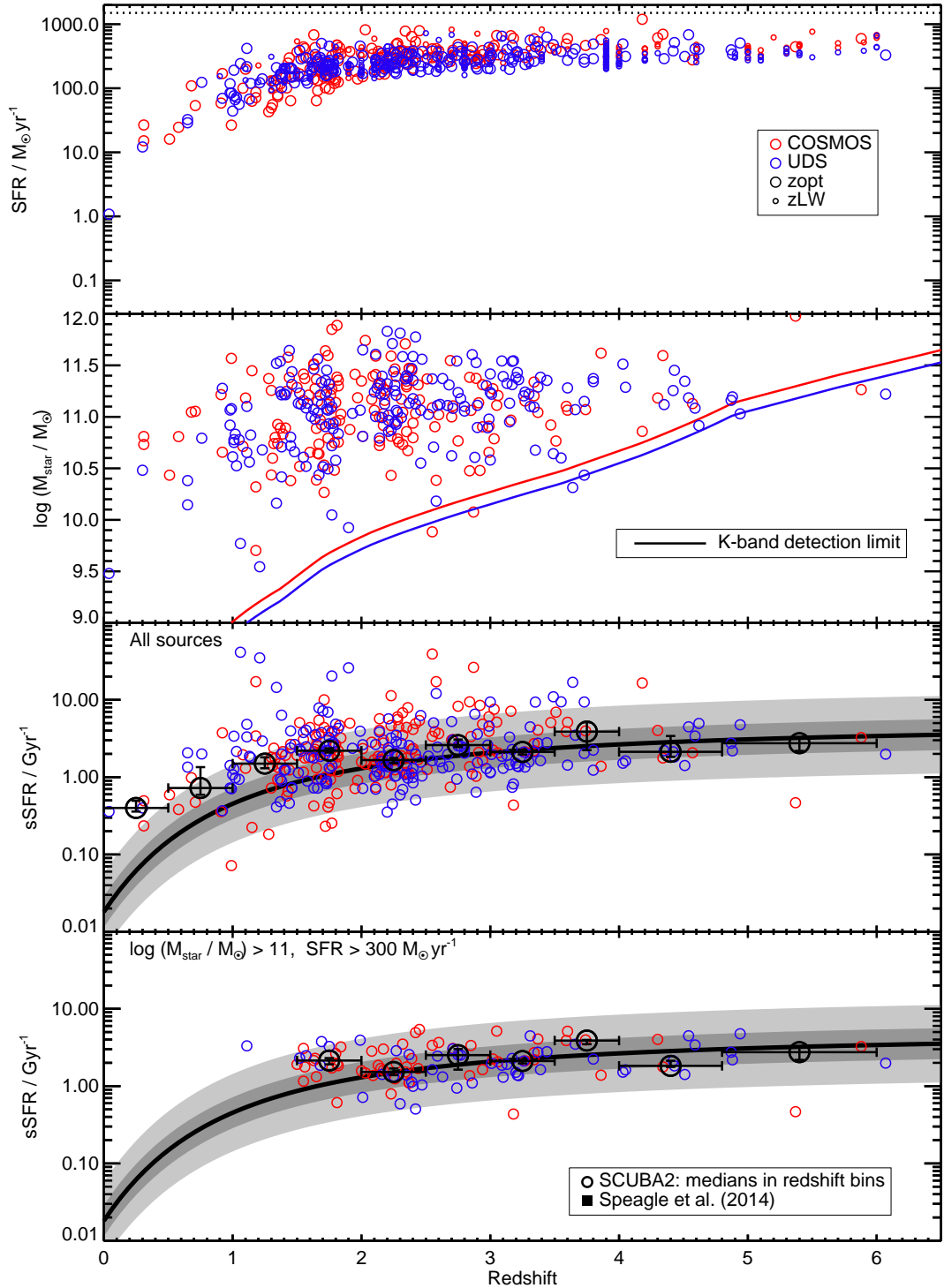}
 \caption{Star formation rates (SFR; {\it first panel}), stellar masses ($\mstar$; {\it second panel}) and specific SFRs ({\it third panel}) of SCUBA-2 sources as a function of redshift. The {\it last panel} also shows the specific SFRs, but including only sources with $\mstar>10^{11}\,\msun$ and $\mbox{SFR}>300\,\msunyr$, as  our survey is sensitive to such objects at all redshifts.  Larger symbols correspond to sources with optical redshifts, whereas smaller symbols to those with only long-wavelength redshifts. The {\it dotted line} on the top panel shows the limit on $\mbox{SFR}=1500\,\msunyr$ above which we do not detect any object. The {\it solid lines} on the second panel show the $3\sigma$ $\mstar$ detection limit corresponding to the $K$-band flux limits from Table~\ref{tab:data}. The {\it solid line} 
 in the two bottom panels represents the main sequence of star-forming galaxies, as measured by \citet{speagle14} plotted for $\log(\mstar/\msun)=11.2$. {\it Light grey} and {\it dark grey} regions represent the $2\sigma$ (0.4 dex) and $1\sigma$ (0.2 dex) scatter in this relation.
 {\it Circles with error bars} on these panels represent the median sSFRs for SCUBA-2 sources in the redshift bins indicated by the horizontal error bars.  The apparent clumps in optical redshifts 
  are due to photometric redshift focusing -- the filters have a given width, so if a spectral feature happens inside one, then it tends to adopt the redshift placing this feature at a similar position with respect to the filter. However, the redshift errors and our adopted redshift bins are larger than this focusing, so this has no effect on our analysis.
}
 \label{fig:sfr}
\end{figure*}

\begin{figure*}
\includegraphics[width=0.8\textwidth]{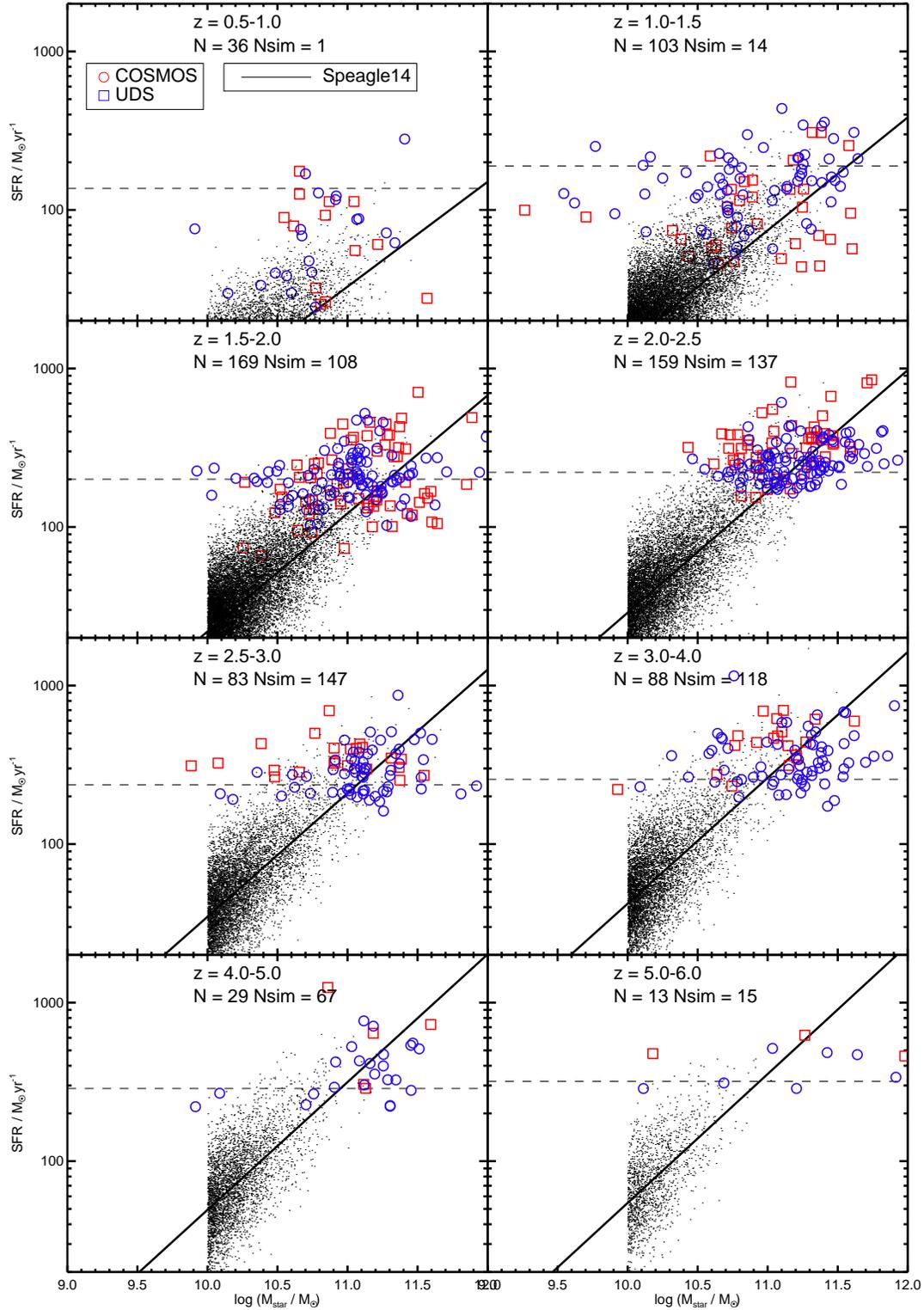}
 \caption{Star formation rates as a function of  stellar mass for the SCUBA-2 sources in the COSMOS ({\it circles}) and UDS ({\it squares}) fields. 
 The {\it solid lines} represent the main sequence of star-forming galaxies at various redshifts, as reported by \citet{speagle14}.
 Most {\smgs} lie on the main sequence. {\it Dots} represent the synthetic main-sequence galaxies distributed according to the mass function of \citet{ilbert13} and \citet{grazian15} and the main sequence reported by \citet{speagle14}, see Section~\ref{sec:sfr}. Their number above the {\smg} SFR threshold ({\it dashed line}), corresponding to 3.5\,mJy, is similar or larger than the number of real {\smgs} {\revone (corrected for completeness)}, which implies that {\smgs}  can be fully explained as the most massive and most highly star-forming main-sequence galaxies, and hence they should not be regarded as a distinct starburst population. We note that the apparent asymmetry in the distribution of synthetic galaxies (enhancement  above the main sequence) is an optical illusion. This can be verified by looking at a narrow mass range, which then shows a perfect symmetry.
}
 \label{fig:sfrm}
\end{figure*}

\begin{figure*}
\includegraphics[width=0.8\textwidth]{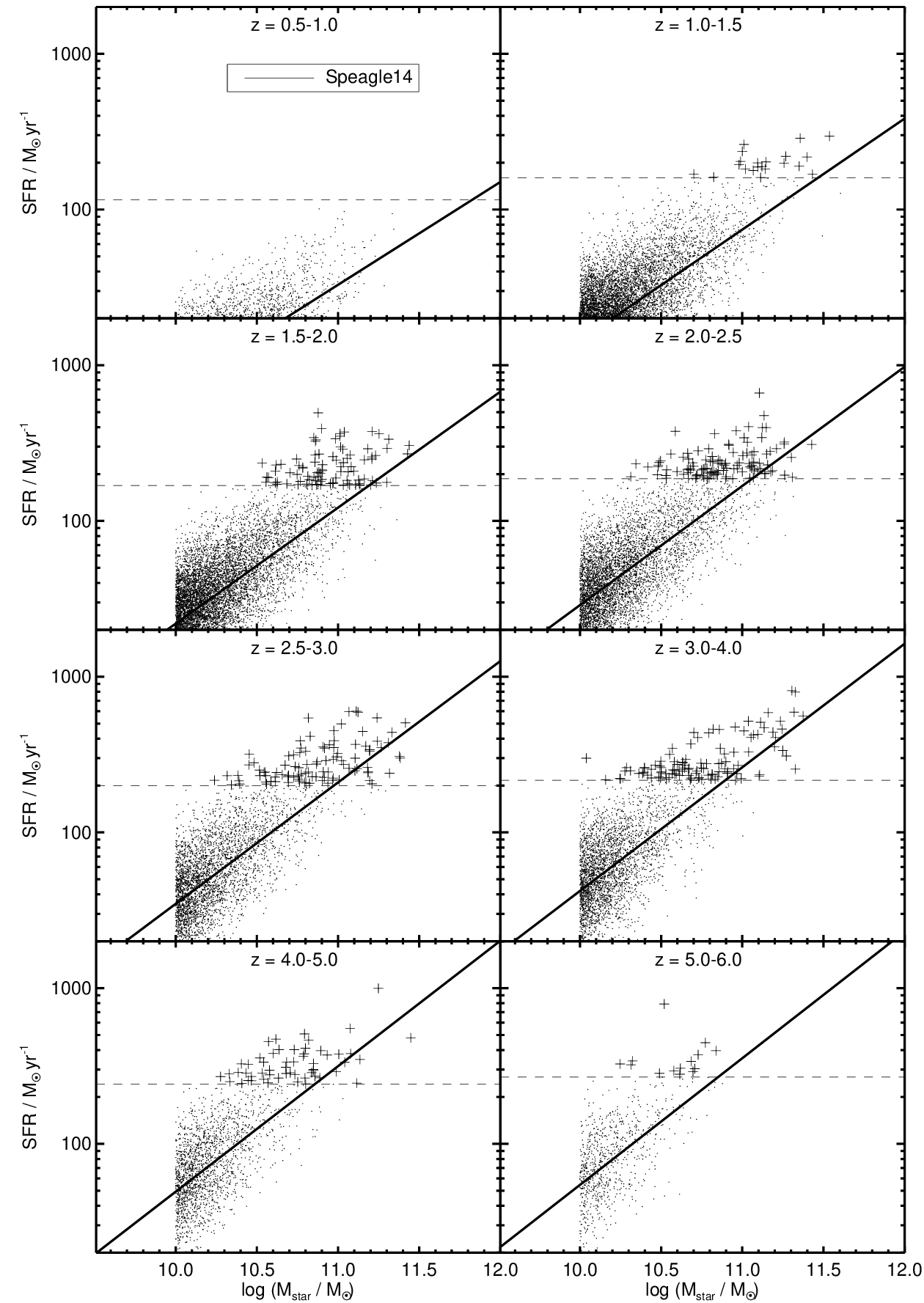}
 \caption{Star formation rates as a function  of  stellar mass for the synthetic galaxies shown on  Fig.~\ref{fig:sfrm}. Those with star formation rates placing them above our $850\,\micron$ survey limit of 3.5\,mJy are marked as {\it crosses}.}
 \label{fig:sfrm_simonly}
\end{figure*}

We estimated SFRs from the fits of the average {\smg}  template \citep[][{\revone dust temperature $\sim39$\,K}]{michalowski10smg} to the $\geq100\,\micron$ photometry assuming either the optical redshift if available, or the long-wavelength redshift (Section~\ref{sec:z}). 
{\revone We integrated the template between $8$ and$1000\,\micron$ and applied the \citet{kennicutt} conversion scaled to the \citet{chabrier03} IMF: $\mbox{SFR}=10^{-10}\times L_{\rm IR}/\lsun$.}
Our data sample the peak of the dust SED, so if we used a hotter SED template \citep[Arp 220;][]{silva98}, then the obtained SFRs would be only $\sim20$--$30$ per cent higher, within the systematic uncertainty of these estimates.
For objects with optical counterparts we estimated stellar masses  from the optical/near-IR SED fits (Section~\ref{sec:z}). 

The resulting  SFRs, stellar masses are given in Tables \ref{tab:IDzCOSMOS} and \ref{tab:IDzUDS} in the appendix.
The SFRs, stellar masses and sSFRs 
are shown as  a function of redshift in Fig.~\ref{fig:sfr}. Table~\ref{tab:sfr} shows median values of these estimates for sources at $z>1$ (excluding discarded optical redshifts, see Section~~\ref{sec:z}). Fig.~\ref{fig:sfrm} shows the SFRs as a function of stellar mass, in comparison with the main sequence of star-forming galaxies 
\citep{speagle14}.

The second panel of Fig.~\ref{fig:sfr} shows that SCUBA-2 sources are very massive galaxies with median masses of $10^{11.15}\,\msun$. in this figure we also show our stellar mass sensitivity limit derived from the $K$-band detection limit (Table~\ref{tab:data}) $k$-corrected to the rest-frame $K$-band luminosity using the average {\smg} template of \citet{michalowski10smg} and using the mass-to-light ratio $\mstar/L_K=0.3\,\msun\lsun^{-1}$. Most of the SCUBA-2 sources are above these limits by an order of magnitude, so our optical/near-IR data is deep enough to ensure the detection of the overwhelming majority of the optical/near-IR counterparts. Hence, our median mass estimate is not biased towards a high value, nor our sSFR estimate is biased towards a low value. These high stellar masses are not directly a result of high SFRs, because, in most cases, $\simeq90$ per cent of the stellar mass  was formed before the currently observed star-formation activity (column 8 of Table~\ref{tab:sfr}). This is consistent with the findings of \citet{dye08} and \citet{michalowski10smg,michalowski12mass}.
{\revtwo 
When modelled, as here, by a single burst, the mean age of earlier star formation is $\simeq1$--$1.5$\,Gyr prior to the epoch of observation. However, we caution that this does not mean that the mass-dominant component was formed in an earlier even more violent short-lived starburst event. Instead, the $\sim$90 per cent of the pre-existing mass could have formed in an extended (several Gyr) period, and indeed could have formed in smaller subcomponents, which subsequently merged.
 }

The lower panels of Fig.~\ref{fig:sfr} and Fig.~\ref{fig:sfrm} show that the  SCUBA-2 sources at $z>2$ (where most of them reside) are fully consistent with the main sequence of star-forming galaxies 
\citep[as quantified  by][]{speagle14}
and form its high-mass end. This is especially highlighted in the fourth panel of this figure, which limits the sample to those with $\mstar>10^{11}\,\msun$ and $\mbox{SFR}>300\,\msunyr$, as our survey is sensitive to such objects  even at $z\simeq5$. In this panel the medians of sSFRs in redshift bins are constant at $z=1$--$6$ and, given the behaviour of the mean sSFR of other galaxies, SCUBA-2 sources stay on the main sequence above $z=1.5$. This is also true for SCUBA-2 sources at $z>4$. This is the first time that a significant sample of {\smgs} at such high redshifts has been studied in relation to the main sequence.  

Even at $1<z<2$ most of the SCUBA-2 sources lie on or close to the main sequence, offset by less than a factor of 2. Only at $z<1$ do the SCUBA-2 sources lie significantly above the main sequence, and correspond to starburst galaxies. At these redshifts our {\submm} flux limit corresponds to a lower luminosity than that at higher redshifts, but the main-sequence normalisation declines even faster from $z\simeq2$ to $z<1$.

Finally, we note that that, for sources that are in fact blends of several sources (Section~4), our SFRs are overestimated, as they include the contribution of other sources, whereas the stellar masses are correct, as long as we identify the correct main contributor to the {\submm} flux. Hence, the true sSFR for these sources are even lower, which makes our conclusion stronger that most of {\smgs} are not above the main sequence.

In order to test whether {\smgs} can indeed be almost exclusively main-sequence galaxies, we considered how many massive main-sequence  galaxies with high SFRs are expected to be located in our $\simeq2.18\,\mbox{deg}^2$ fields, given what we know about the galaxy stellar mass function and SFRs of star-forming galaxies at a given redshift. To estimate the expected number density of such 
objects we used the mass function of \citet{ilbert13} at $z<4$ and of \citet{grazian15} at $z>4$. For each redshift bin shown in Fig.~\ref{fig:sfrm} we multiplied the integral of the corresponding mass function between $\log(\mstar/\msun)=10$--$12$ (the range spanned by {\smgs}) with the volume probed by our survey within this redshift bin to obtain the total number of star-forming galaxies in this mass range expected in our fields. Their masses were chosen randomly out of the mass function, so that the resulting mass distribution matches the measured mass function. To each of these synthetic galaxies  we assigned an SFR based on the main sequence at that redshift \citep{speagle14} and scattered them randomly by a number drawn from a Gaussian distribution with a standard deviation of 0.2\,dex \citep[the width of the main sequence;][]{speagle14}. These synthetic main-sequence galaxies are shown as dots in Fig.~\ref{fig:sfrm} and the number of them above the SFR cut corresponding to {\smgs} (dashed line) is shown on each panel as `Nsim' (these most-star-forming synthetic galaxies are clearly marked as plus signs in Fig.~\ref{fig:sfrm_simonly}). 

Between $z \simeq 1$ and $z \simeq 4$ the number of predicted and observed bright submm 
galaxies is in very good agreement (to within a factor of 2) given the relative simplicity of this calculation.
Indeed the predicted number is always larger that what is actually observed, particularly so at $z > 4$, and 
so given current data on the evolution of the galaxy mass function and the main sequence, there is clearly no problem explaining the prevalence of submm galaxies at all redshifts.

There are some obvious reasons that this calculation may overpredict somewhat the observed number
of submm galaxies at the highest redshifts. Given the small number statistics at $z > 4$ redshift errors 
may be important, and in addition our completeness may be poorer than estimated. However, it is 
equally likely that the predicted number of massive star-forming galaxies may be in error at the 
highest redshifts, given our current limited knowledge of the form of the galaxy 
stellar mass function at $z > 4$ (the high-mass end being particularly vulnerable to systematic errors
such as Eddington bias).

Nonetheless, these calculations, as illustrated in Fig.~\ref{fig:sfrm} and Fig.~\ref{fig:sfrm_simonly}, clearly demonstrate that the 
observed properties of luminous high-redshift submm galaxies arise naturally from the evolving 
main sequence of normal star-forming objects, once the selection function inherent in submm 
surveys is taken into account.

\section{Discussion: extreme star formation in the Universe}
\label{sec:discuss}
 
 \begin{figure}
\includegraphics[width=0.45\textwidth]{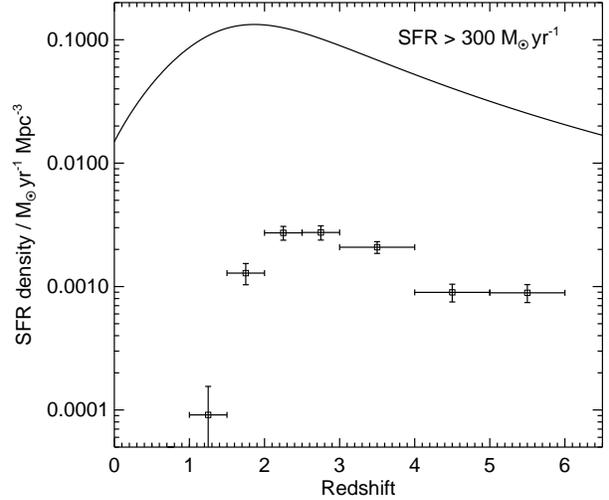}
 \caption{The comoving star formation rate density contributed by {\smgs} with $\mbox{SFR}\geq300\,\msunyr$ ({\it squares}; our survey is sensitive to such objects at all redshifts, see Fig.~\ref{fig:sfr}). The {\it Solid line} indicates a recent determination of the total SFR density \citep{madau14}.
}
 \label{fig:sfrden}
\end{figure}

\begin{figure}
\includegraphics[width=0.45\textwidth]{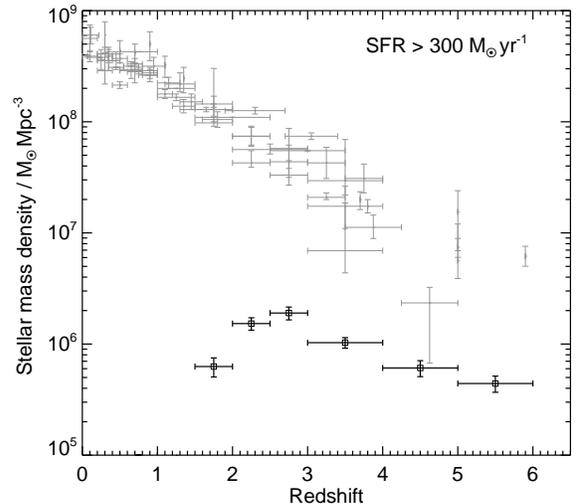}
 \caption{The comoving stellar mass density contributed by {\smgs} with $\mbox{SFR}\geq300\,\msunyr$ ({\it squares}; our survey is sensitive to such objects at all redshifts, see Fig.~\ref{fig:sfr}).  {\it Grey points} represent the total stellar mass density compiled in \citet{madau14} .
}
 \label{fig:msden}
\end{figure}

\begin{figure}
\includegraphics[width=0.45\textwidth]{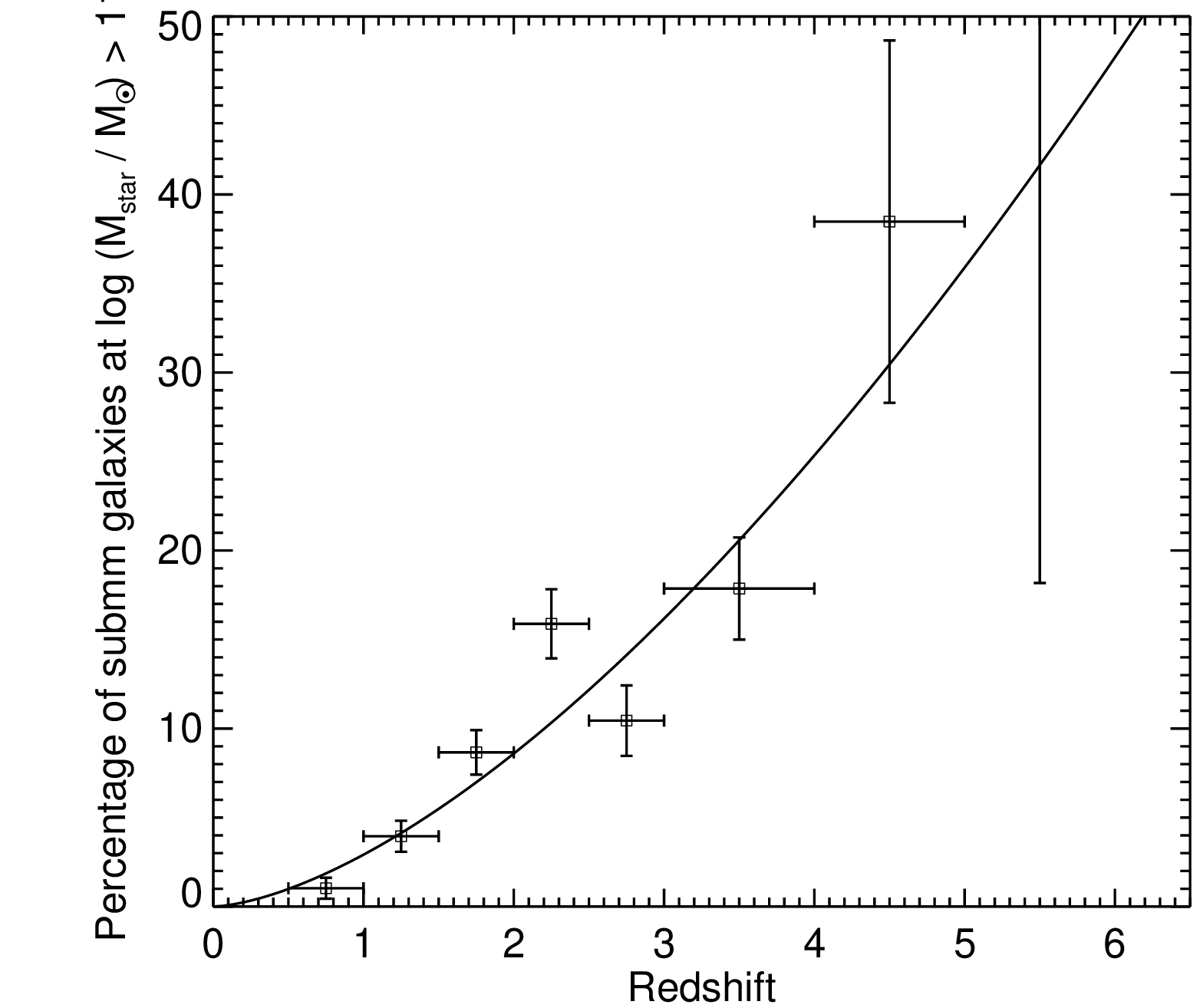}
 \caption{The fraction of galaxies with masses above $\log(\mstar/\msun)=11$ that are {\smgs} ({\it squares}), derived from the comparison of the number density of {\smgs} and the integral of the total mass function of star-forming galaxies at a given redshift \citep{ilbert13,grazian15,caputi15} above this stellar mass. The {\it Solid line} is a power-law fit of the form of $(2.9\pm0.4)\times z^{1.56\pm0.16}$.}
 \label{fig:fracsmg}
\end{figure}

\begin{table*}
\caption{Comoving number density, SFR density, and stellar mass density of {\smgs} and fraction of galaxies with $\log(\mstar/\msun)\geq11$ which are {\smgs} based on our $\geq4\sigma$ sample with 1.1\,mm coverage.}
\label{tab:fracsmg}
\begin{center}
\begin{tabular}{ccccc}
\hline\hline
$z$ & $n_{\rm den}$ & $\rho_{\rm SFR}$ & $\rho_{\mstar}$ &  $\mbox{frac}_{\rm SMG}$ \\
    &   ($10^{-6}\mbox{Mpc}^{-3}$) & ($10^{-3}\msunyr\,\mbox{Mpc}^{-3}$) & ($10^{6}\msun\,\mbox{Mpc}^{-3}$) &  (\%)    \\
\hline
0.5--1.0 & $0.00\pm0.00$ & $0.00\pm0.00$ & $0.00\pm0.00$ & $1.0\pm0.6$ \\
1.0--1.5 & $0.37\pm0.26$ & $0.09\pm0.06$ & $0.03\pm0.02$ & $3.9\pm0.9$ \\
1.5--2.0 & $4.02\pm0.78$ & $1.29\pm0.25$ & $0.63\pm0.12$ & $8.7\pm1.2$ \\
2.0--2.5 & $8.83\pm1.13$ & $2.72\pm0.35$ & $1.53\pm0.20$ & $15.9\pm1.9$ \\
2.5--3.0 & $8.27\pm1.10$ & $2.74\pm0.36$ & $1.90\pm0.25$ & $10.4\pm2.0$ \\
3.0--4.0 & $6.42\pm0.70$ & $2.08\pm0.23$ & $1.03\pm0.11$ & $17.9\pm2.9$ \\
4.0--5.0 & $3.14\pm0.51$ & $0.90\pm0.15$ & $0.61\pm0.10$ & $38.5\pm10.2$ \\
5.0--6.0 & $3.41\pm0.57$ & $0.89\pm0.15$ & $0.44\pm0.07$ & $60.8\pm42.6$ \\
\hline
\end{tabular}
\end{center}
\end{table*}

\subsection{Main-sequence nature and the maximum SFR}
     
 \citet{koprowski16} showed that {\smgs} in deep SCUBA-2 fields are located on the main sequence, and now we have obtained a similar result for a brighter sample from shallower but larger fields. 
This is incompatible with the frequently assumed picture that {\smgs} are unusually powerful starbursts, significantly different from the general star-forming galaxy population. Instead, we have shown that {\submm} surveys simply (and inevitably) select the most massive (and hence most star-forming) galaxies out of the main-sequence population. This suggests that most {\smgs} are not fuelled by extreme, transitory
event such as a major merger (which would move them above the main sequence), but instead represent the final stages (shortly prior to quenching) of a long and (on average) fairly smooth, ascent up the main sequence. This interpretation is supported by recent simulations showing that all properties of {\smgs} can be explained by a sustained gas inflow, rather than by major mergers \citep{narayanan15}.

It is clear however, that {\it some} {\smgs} are powered by major mergers.
CO and H$\alpha$ observations revealed that in roughly half of the {\smgs} gas is distributed in multiple components and in the remaining half the gas distribution is compact  \citep{tacconi08,engel10,alaghbandzadeh12}. This is consistent with a major merger scenario, but also with a clumpy disc scenario if the separation is not too large. On the other hand, near-IR \citep{targett11,targett13,wiklind14,chen15}, kinematic \citep{swinbank11,hodge12,menendezdelmestre13} and resolved dust/gas studies \citep{bothwell10,hodge15,hodge16} of {\smgs} indeed reveals that some of them are large, clumpy disc galaxies, sometimes with potential merger signatures.

Due to our large and well-defined bright SCUBA-2 sample, this is the first time that it has proved possible 
to properly investigate the position of {\smgs} at $z>4$ relative to the main sequence. As demonstrated in Figs~\ref{fig:sfr} and \ref{fig:sfrm}, even at such high redshifts, {\smgs} are consistent on average with the main sequence. Hence, these galaxies represent the most powerful star-forming galaxies at these early epochs, but again are likely not powered by any unusual/extreme events.

It is instructive to investigate whether there is a limit to the SFR of {\smgs}. We have not detected any source above $\mbox{SFR}=1500\,\msunyr$ (dotted line on the top panel of Fig.~\ref{fig:sfr}). This is because the sources in our sample do not exceed the $850\,\micron$ flux of $17$\,mJy. This implies an upper limit on the number density of such extreme sources of $<0.023\,\mbox{deg}^2$ (95 per cent confidence). This was calculated as 1/20 of a number density if there was one galaxy per $2.17\,\mbox{deg}^2$ field (the area of the combined UDS and COSMOS SCUBA-2 fields) and can be confirmed by generating random positions in a large area (e.g.~100\,deg$^2$) and checking that, at this surface density, 95 per cent of random $2.17\,\mbox{deg}^2$ fields contain no sources.  Sources more active than  $\mbox{SFR}=1500\,\msunyr$ have been confirmed in the past \citep{capak08,capak11,daddi09,michalowski10smg4, riechers10,riechers13,hezaveh13}, but usually they were just single objects in given fields, so the estimate of their number density is difficult. 

We note that our SFR cutoff value is higher than the maximum of $1000\,\msunyr$ proposed by \citet{karim13} based on the lack of $>9$\,mJy sources in the ALMA follow-up of LESS sources. However, their smaller parent single-dish sample contained only a few such sources, so the conclusion presented here is more robust.  Indeed, $>9$\,mJy interferometric sources were detected by both the Submillimeter Array \citep[SMA;][]{younger07, younger09b,barger12} and ALMA \citep{simpson15,simpson15b}.  
On the other hand \citet{barger14} found a turn-down in the SFR distribution function above $\sim1100\,\msunyr$ (after the conversion to our adopted \citet{chabrier03} IMF), which implies that such sources become increasingly rare, which is compatible with our cut-off value.

\subsection{{\Smgs} without IDs}

We have not found any IDs for around a third of SCUBA-2 sources (Table~\ref{tab:succ}). They can be divided into three categories: (1) spurious or flux-boosted {\submm} sources, (2) blends of several {\submm} sources out of which none is bright enough for our ID method to work and (3) high redshift sources, with too low radio and mid-IR fluxes. Our ID method is likely to miss spurious sources (they would be very unlikely to yield  IDs), and, as demonstrated in Fig.~\ref{fig:ALMAID}, would not identify many faint {\submm} sources,
especially at high redshift. Hence, our ID catalogue should reflect a relatively clean {\submm}-flux-limited sample (removing problematic categories 1 and 2), but may underrepresent  very high redshift sources, which are likely not to be IDed either.
Indeed, the ID completeness of the entire sample ($\sim66$ per cent) is lower than the ID completeness in the ALMA subsample ($\sim86$ per cent). This is likely because the ALMA subsample is brighter \citep[fig.~2 of][]{simpson15b}, so it contains less spurious sources, for which our method would (correctly) return no ID.

We can estimate the fraction of sources in these categories based on our ID fraction (Table~\ref{tab:succ}) and the ALMA training sample \citep[][and Section~\ref{sec:id}]{simpson15b}.
In Section~\ref{sec:id} we showed that for four out of  29 ALMA-observed SCUBA-2 sources our ID method misses  dominant ALMA sources. One of them is not covered by the $24\,\mu$m imaging, but the lack of IDs for other three ($\sim10$ per cent) indicates that they can  be at very high redshifts. They are unlikely to be spurious sources, as the fluxes are confirmed by ALMA at the $\sim4$--$8$\,mJy level. Two of them are not detected by {\it Herschel} implying a long-wavelength redshift of $>3$ and $>4$ (Section~\ref{sec:z}). The third has a significant {\it Herschel} signal, but there are two very strong $24\,\mu$m sources nearby complicating the photometry. In any case, some other SCUBA-2 sources with no IDs (for which we do not have ALMA data) may also belong to the high-$z$ category. This can be tested by high-resolution {\submm} interferometry and subsequent CO redshift search.

Some sources can be affected by blending. For three out of 29 ALMA-observed SCUBA-2 sources ($\sim10$ per cent) the brightest ALMA component is fainter than half  of the SCUBA-2 flux. Additionally, for two SCUBA-2 sources there are no ALMA counterparts. This means that for $\sim17$ per cent of the SCUBA-2 sample the true {\submm} flux may be twice lower than measured, making it difficult to find IDs.
We also note that multiplicity should not influence our long-wavelength estimates, because if given sources are blended at the JCMT/SCUBA-2 resolution, then they are blended at the {\it Herschel}/SPIRE resolution. Hence, far-IR colours of the main contributor to the {\submm} flux are not significantly affected, unless the sources are at significantly different redshifts. Hence, sources with no IDs with $z_{\rm LW}\sim2$ are likely blends of galaxies at that redshift, whereas those with no IDs and  $z_{\rm LW}\gtrsim4$ are likely truly at these high redshifts, as blending should not result in an artificially high   $z_{\rm LW}$.

\subsection{{\Smgs} in cosmological context}

Fig.~\ref{fig:sfrden} and Fig.~\ref{fig:msden} show the contribution of our {\smgs} to the cosmic SFR and stellar mass densities, respectively. 
{\revone The values are shown in Table~\ref{tab:fracsmg}.}
{\revone To calculate the volume of each redshift bin we assumed the combined area of the COSMOS and UDS SCUBA-2 imaging with $1.1$\,mm coverage of $1.15\,\mbox{deg}^2$. For each source we used our best redshift, either optical, or long-wavelength if optical was not available or rejected. 
Completeness corrections have been applied as described in \citet{geach17}.
}

Bright {\smgs}, as studied here, contribute $\simeq2$--$4$ per cent of the SFR density at $z=2$--$6$, and $\simeq3$ per cent to the stellar mass density at $z=2$--$4$, rising to $\simeq 10$ per cent at $z=4$--$6$.
Deeper mm/{\submm} surveys with SCUBA-2 {\revone \citep[e.g.][]{casey13,barger14,coppin15,bourne17}} and ALMA
\citep{dunlop17} show that fainter dusty star-forming galaxies contribute the vast majority of cosmic star formation rate density at $z  = 1$--$3$.

Finally, in Fig.~\ref{fig:fracsmg} and Table~\ref{tab:fracsmg} we show the fraction of star-forming galaxies above $\log(\mstar/\msun)=11$ that are {\smgs}, calculated by dividing the number density of {\smgs} with $\log(\mstar/\msun)\geq11$ in a given redshift bin by the integral of the mass function \citep{ilbert13,grazian15,caputi15} above that mass.  The power-law fit to this fraction results in the following dependence: $(2.9\pm0.4)\times z^{1.56\pm0.16}$.  The fraction of {\smgs} increases with redshift and reaches $\simeq30$ per cent at $z=4$. This is because our selection function is nearly flat with redshift, whereas the normalisation of the main sequence is increasing, so the fraction of massive galaxies that should be detectable above our SFR-limited flux-density limit is expected to increase
(albeit the total number density of such massive galaxies obviously rapidly declines 
with increasing redshift).

\section{Conclusions}
\label{sec:conclusion}

We have conducted an analysis of nearly $2000$ submm sources
detected in the $\simeq 2\,\mbox{deg}^2$ $850$-$\micron$ imaging of the COSMOS and UDS fields
obtained with SCUBA-2 on the JCMT as part of the SCUBA-2 Cosmology Legacy Survey. 
This unique data set represents the largest homogeneous sample of 850-$\mu$m-selected sources assembled to date, and we have exploited this sample, along with the rich
multiwavelength supporting data in these fields to shed new light on the physical properties and 
cosmological evolution of bright ($S_{850} \ge 4$\,mJy) submm-selected galaxies.

We have completed the galaxy identification process for all 850-$\mu$m sources selected with S/N~$\ge$~3.5, but focus our scientific analysis on a high-quality subsample of 651 sources selected with S/N~$\ge$~4 and complete multiwavelength coverage extending to include 1.1-mm imaging. We have 
checked the
reliability of our identifications, and the robustness of the SCUBA-2 
fluxes, by revisiting the results of recent ALMA follow-up of a subset
of the brightest sources in our sample. This shows that our identification method has a completeness
of $\simeq 86$ per cent with a reliability of $\simeq 92$ per cent, and that only $\simeq 15$--$20$ per cent of sources
are significantly affected by multiplicity. For completeness, we have also shown 
that the impact of source blending on
the 850-$\mu$m source counts as determined with SCUBA-2 is modest; scaling the single-dish
fluxes by $\simeq 0.9$ reproduces the ALMA source counts. 

The optical/near-IR/mid-IR data, coupled at longer wavelengths with the {\it Herschel}+SCUBA-2+AzTEC
photometry, have enabled us to estimate the redshifts ($z$) and star formation rates (SFR) of {\it all} 
sources in our {\revone entire} sample, and stellar masses ($\mstar$) for the $\simeq 75$ per cent of 
sources with optical/near-IR galaxy identifications.

For our $4\sigma$ sample {\revone with 1.1\,mm coverage} we find median values of $z = 2.40^{+0.10}_{-0.04}$, $\mbox{SFR} = 287\pm6\,\msunyr$ and $\log(\mstar/\msun) = 11.12\pm0.02$ {\revtwo (the latter for 349/651 sources with optical identifications)}, and we have shown that these properties clearly locate
bright {\submm} galaxies on the high-mass end of the `main sequence' of star-forming galaxies
out to $z \simeq 6$,
suggesting that major mergers are not a dominant driver of the high-redshift submm-selected population. 
We have also shown that the number densities of these high-mass
main-sequence galaxies are consistent with recent determinations of the evolving
galaxy stellar mass function, and have calculated the contributions of these most luminous star-forming
main-sequence galaxies to cosmic star formation rate density and cosmic stellar mass density as a function
of redshift. 

We conclude that the {\submm} galaxy population is essentially 
as expected (both in terms of evolving comoving number density, and with regard to 
inferred physical properties), albeit reproducing the evolution of the main sequence of star-forming galaxies 
remains a challenge for theoretical models/simulations.

\section*{Acknowledgements}

We  thank Joanna Baradziej, Ian Smail, James Simpson, and our anonymous referee
 for comments and suggestions.

MJM acknowledges the support of the UK Science and Technology Facilities Council (STFC), British Council Researcher Links Travel Grant and the hospitality at the Instituto Nacional de Astrof\'{i}sica, \'{O}ptica y Electr\'{o}nica.
JSD acknowledges the support of the European Research Council via the award of an Advanced Grant, and the contribution of the EC FP7 SPACE project ASTRODEEP (Ref. No: 312725). 
MPK acknowledges the STFC Studentship Enhancement Programme grant and the Carnegie Trust Research Incentive Grant (PI: Micha{\l}owski).

The James Clerk Maxwell Telescope has historically been operated by the Joint Astronomy Centre on behalf of the Science and Technology Facilities Council of the United Kingdom, the National Research Council of Canada and the Netherlands Organisation for Scientific Research. Additional funds for the construction of SCUBA-2 were provided by the Canada Foundation for Innovation. 

This work is based on data products from observations made with ESO Telescopes at the La Silla Paranal Observatory  as  part  of  programme  ID  179.A-2005,  using  data products  produced  by  TERAPIX  and  the  Cambridge  Astronomy  Survey  Unit  on  behalf  of  the  UltraVISTA  consortium. 

This study 
was based in part on observations obtained with MegaPrime/MegaCam, a joint project of CFHT and CEA/DAPNIA, at the Canada-France-Hawaii Telescope (CFHT) which is operated by the National Research Council (NRC) of Canada, the Institut National  des  Science  de  l'Univers  of  the  Centre  National de  la  Recherche  Scientifique  (CNRS)  of  France,  and  the University  of  Hawaii.  This  work  is  based  in  part  on  data products  produced  at  TERAPIX  and  the  Canadian  Astronomy Data Centre as part of the Canada-France-Hawaii Telescope  Legacy  Survey,  a  collaborative  project  of  NRC and  CNRS. 

The National Radio Astronomy Observatory is a facility of the National Science Foundation operated under cooperative agreement by Associated Universities, Inc. 

This work is based in part on observations made with the Spitzer Space Telescope, which is operated by the Jet Propulsion Laboratory, California Institute of Technology under a contract with NASA. 

 This research has made use of data from HerMES project (http://hermes.sussex.ac.uk/). HerMES is a Herschel Key Programme utilising Guaranteed Time from the SPIRE instrument team, ESAC scientists and a mission scientist. 
 The HerMES data was accessed through the Herschel Database in Marseille (HeDaM - http://hedam.lam.fr) operated by CeSAM and hosted by the Laboratoire d'Astrophysique de Marseille. 
PACS has been developed by a consortium of institutes led by MPE (Germany) and including UVIE (Austria); KU Leuven, CSL, IMEC (Belgium); CEA, LAM (France); MPIA (Germany); INAF-IFSI/OAA/OAP/OAT, LENS, SISSA (Italy); IAC (Spain). This development has been supported by the funding agencies BMVIT (Austria), ESA-PRODEX (Belgium), CEA/CNES (France), DLR (Germany), ASI/INAF (Italy), and CICYT/MCYT (Spain). 
SPIRE has been developed by a consortium of institutes led by Cardiff University (UK) and including Univ. Lethbridge (Canada); NAOC (China); CEA, LAM (France); IFSI, Univ. Padua (Italy); IAC (Spain); Stockholm Observatory (Sweden); Imperial College London, RAL, UCL-MSSL, UKATC, Univ. Sussex (UK); and Caltech, JPL, NHSC, Univ. Colorado (USA). This development has been supported by national funding agencies: CSA (Canada); NAOC (China); CEA, CNES, CNRS (France); ASI (Italy); MCINN (Spain); SNSB (Sweden); STFC (UK); and NASA (USA). 

This work is based on observations taken by the 3D-HST Treasury Program (GO 12177 and 12328) with the NASA/ESA HST, which is operated by the Association of Universities for Research in Astronomy, Inc., under NASA contract NAS5-26555. 
Based on data obtained with the European Southern Observatory Very Large Telescope, Paranal, Chile, under Large Program 185.A-0791, and made available by the VUDS team at the CESAM data center, Laboratoire d'Astrophysique de Marseille, France. The HST data matched to the VUDS-DR1 are described in \citet{grogin11} and \citet{koekemoer11} for CANDELS and include data from the ERS \citep{windhorst11}.  
This paper uses data from the VIMOS Public Extragalactic Redshift Survey (VIPERS). VIPERS has been performed using the ESO Very Large Telescope, under the "Large Programme" 182.A-0886. The participating institutions and funding agencies are listed at \url{http://vipers.inaf.it} 
Based    on    zCOSMOS    observations    carried    out    using    the    Very    Large    Telescope    at    the    ESO    Paranal    Observatory    under    Programme    ID: LP175.A-0839 

 This research has made use of the Tool for OPerations on Catalogues And Tables \citep[TOPCAT;][]{topcat}: \url{www.starlink.ac.uk/topcat/ };
 SAOImage DS9, developed by Smithsonian Astrophysical Observatory \citep{ds9}; SExtractor: Software for source extraction \citep{sextractor},
 and NASA's Astrophysics Data System Bibliographic Services.




\appendix

\section{Online tables}
\label{sec:fig}

\clearpage

\onecolumn


\renewcommand{\tabcolsep}{1.5ex}
\begin{tiny}
%
\begin{longtable}{rllccccllccccllcccc}
\caption{Radio, $24\,\mu$m and $8\,\mu$m identifications of the JCMT/SCUBA2 objects in the COSMOS field. This table is available in its entirety in the online version.}
\label{tab:IDCOSMOS}\\
\hline\hline
No. &  RA$_{1.4}$ & DEC$_{1.4}$ & F$_{1.4}$ & E$_{1.4}$ & Sep & $p$ &  RA$_{24}$ & DEC$_{24}$ & F$_{24}$ & E$_{24}$ & Sep & $p$ &  RA$_{8}$ & DEC$_{8}$ & F$_{8}$ & E$_{8}$ & Sep & $p$ \\
    &  (deg)      & (deg)     & ($\mu$Jy)& ($\mu$Jy)& ($''$)     &     &  (deg)      & (deg)     & ($\mu$Jy)& (mJy)    &  ($''$)    &     &  (deg)      & (deg)     & ($\mu$Jy)& (mJy)    &  ($''$) &          \\
\hline
1 & $\cdots$ & $\cdots$ &$\cdots$ &$\cdots$ &$\cdots$ &$\cdots$ & $\cdots$ & $\cdots$ &$\cdots$ &$\cdots$ &$\cdots$ &$\cdots$ & $\cdots$ & $\cdots$ &$\cdots$ &$\cdots$ &$\cdots$ &$\cdots$ \\
2 & 149.65820 &2.2357170 &140 &\phantom{1}10 &2.3 &0.0035 &149.65810 &2.2357270 &290 &\phantom{1}10 &2.3 &0.0220 &149.65820 &2.2356580 &\phantom{1}24 &\phantom{1}\phantom{1}2 &2.1 &0.0620 \\
3 & 150.03340 &2.4367110 &\phantom{1}80 &\phantom{1}10 &1.4 &0.0027 &150.03340 &2.4365270 &280 &\phantom{1}10 &1.8 &0.0150 &$\cdots$ & $\cdots$ &$\cdots$ &$\cdots$ &$\cdots$ &$\cdots$ \\
$.$  & $\cdots$ & $\cdots$ &$\cdots$ &$\cdots$ &$\cdots$ &$\cdots$ & $\cdots$ & $\cdots$ &$\cdots$ &$\cdots$ &$\cdots$ &$\cdots$ & 150.03340 &2.4359700 &\phantom{1}59 &\phantom{1}\phantom{1}2 &3.4 &0.0610 \\
4 & $\cdots$ & $\cdots$ &$\cdots$ &$\cdots$ &$\cdots$ &$\cdots$ & 149.92860 &2.4935800 &\phantom{1}50 &\phantom{1}10 &0.6 &0.0130 &149.92850 &2.4939570 &\phantom{1}14 &\phantom{1}\phantom{1}2 &1.8 &0.0750 \\
5 & 150.09990 &2.2972110 &190 &\phantom{1}50 &0.7 &0.0003 &150.09990 &2.2973210 &160 &\phantom{1}10 &0.5 &0.0040 &150.10010 &2.2971450 &\phantom{1}35 &\phantom{1}\phantom{1}2 &1.0 &0.0120 \\
\hline
\end{longtable}

\begin{longtable}{rllccccllccccllcccc}
\caption{Radio, $24\,\mu$m and $8\,\mu$m identifications of the JCMT/SCUBA2 objects in the UDS field. This table is available in its entirety in the online version.}
\label{tab:IDUDS}\\
\hline\hline
No. &  RA$_{1.4}$ & DEC$_{1.4}$ & F$_{1.4}$ & E$_{1.4}$ & Sep & $p$ &  RA$_{24}$ & DEC$_{24}$ & F$_{24}$ & E$_{24}$ & Sep & $p$ &  RA$_{8}$ & DEC$_{8}$ & F$_{8}$ & E$_{8}$ & Sep & $p$ \\
    &  (deg)      & (deg)     & ($\mu$Jy)& ($\mu$Jy)& ($''$)     &     &  (deg)      & (deg)     & ($\mu$Jy)& (mJy)    &  ($''$)    &     &  (deg)      & (deg)     & ($\mu$Jy)& (mJy)    &  ($''$) &          \\
\hline
1 & 34.62779 &-5.5254170 &271 &\phantom{1}30 &1.7 &0.0017 &34.62744 &-5.5255270 &416 &\phantom{1}19 &3.0 &0.0190 &$\cdots$ & $\cdots$ &$\cdots$ &$\cdots$ &$\cdots$ &$\cdots$ \\
1 & $\cdots$ & $\cdots$ &$\cdots$ &$\cdots$ &$\cdots$ &$\cdots$ & 34.62838 &-5.5247980 &\phantom{1}40 &\phantom{1}\phantom{1}4 &1.6 &0.0570 &$\cdots$ & $\cdots$ &$\cdots$ &$\cdots$ &$\cdots$ &$\cdots$ \\
2 & 34.60079 &-5.3822500 &144 &\phantom{1}36 &2.3 &0.0066 &34.60092 &-5.3824720 &121 &\phantom{1}\phantom{1}4 &1.7 &0.0360 &34.60068 &-5.3822010 &\phantom{1}32 &\phantom{1}\phantom{1}3 &2.7 &0.0690 \\
3 & 34.83821 &-4.9476670 &\phantom{1}65 &\phantom{1}21 &1.0 &0.0044 &34.83805 &-4.9481440 &\phantom{1}67 &\phantom{1}\phantom{1}8 &1.1 &0.0280 &34.83807 &-4.9474830 &\phantom{1}23 &\phantom{1}\phantom{1}2 &1.7 &0.0470 \\
4 & 34.19967 &-5.0249170 &\phantom{1}68 &\phantom{1}22 &2.6 &0.0199 &34.19946 &-5.0249320 &187 &\phantom{1}\phantom{1}3 &3.3 &0.0630 &34.19976 &-5.0249510 &\phantom{1}46 &\phantom{1}\phantom{1}4 &2.2 &0.0360 \\
5 & 34.35725 &-5.4281950 &103 &\phantom{1}19 &4.7 &0.0306 &34.35717 &-5.4280670 &444 &\phantom{1}\phantom{1}7 &4.4 &0.0310 &34.35736 &-5.4282440 &\phantom{1}82 &\phantom{1}\phantom{1}7 &4.9 &0.0740 \\
\hline
\end{longtable}

\begin{longtable}{rccccccc}
\caption{Long-wavelength fluxes of the JCMT/SCUBA2 objects in the COSMOS field. This table is available in its entirety in the online version.}
\label{tab:IDFlongCOSMOS}\\
\hline\hline
No. &  F100 & F160 & F250 & F350 & F500 & F850 & F11   \\
    &  (mJy) & (mJy) & (mJy) & (mJy) & (mJy) & (mJy) & (mJy) \\
\hline
1 & $\cdots$ & $\cdots$ & \phantom{1}11.07 $\pm$ \phantom{1}6.02 & \phantom{1}17.93 $\pm$ \phantom{1}6.70 & \phantom{1}18.85 $\pm$ \phantom{1}7.27 & 12.9 $\pm$ 0.9 & 8.85 $\pm$ 1.09 \\ 
2 & \phantom{1}\phantom{1}1.14 $\pm$ 0.79 & \phantom{1}\phantom{1}2.83 $\pm$ \phantom{1}1.95 & \phantom{1}26.32 $\pm$ \phantom{1}6.63 & \phantom{1}39.79 $\pm$ \phantom{1}8.03 & \phantom{1}36.77 $\pm$ \phantom{1}7.21 & 13.2 $\pm$ 1.0 & 8.69 $\pm$ 1.31 \\ 
3 & \phantom{1}\phantom{1}3.43 $\pm$ 0.81 & \phantom{1}\phantom{1}9.02 $\pm$ \phantom{1}1.93 & \phantom{1}23.51 $\pm$ \phantom{1}5.99 & \phantom{1}31.42 $\pm$ \phantom{1}6.60 & \phantom{1}29.02 $\pm$ \phantom{1}7.29 & 15.4 $\pm$ 1.4 & 9.81 $\pm$ 1.36 \\ 
$.$  & $\cdots$ & $\cdots$ & \phantom{1}\phantom{1}0.00 $\pm$ \phantom{1}6.68 & \phantom{1}\phantom{1}0.00 $\pm$ \phantom{1}7.36 & \phantom{1}\phantom{1}0.00 $\pm$ \phantom{1}7.95 & 15.4 $\pm$ 1.4 & 9.81 $\pm$ 1.36 \\ 
4 & \phantom{1}\phantom{1}0.70 $\pm$ 0.79 & \phantom{1}\phantom{1}2.95 $\pm$ \phantom{1}1.58 & \phantom{1}17.92 $\pm$ \phantom{1}6.00 & \phantom{1}27.98 $\pm$ \phantom{1}6.68 & \phantom{1}25.44 $\pm$ \phantom{1}7.08 & 16.7 $\pm$ 1.5 & 10.89 $\pm$ 1.30 \\ 
5 & \phantom{1}\phantom{1}2.75 $\pm$ 0.78 & \phantom{1}\phantom{1}0.67 $\pm$ \phantom{1}1.59 & \phantom{1}15.07 $\pm$ \phantom{1}6.19 & \phantom{1}29.50 $\pm$ \phantom{1}7.47 & \phantom{1}26.84 $\pm$ \phantom{1}7.87 & \phantom{1}9.6 $\pm$ 0.9 & 3.09 $\pm$ 1.18 \\ 
\hline
\end{longtable}

\begin{longtable}{rccccccc}
\caption{Long-wavelength fluxes of the JCMT/SCUBA2 objects in the UDS field. This table is available in its entirety in the online version.}
\label{tab:IDFlongUDS}\\
\hline\hline
No. &  F100 & F160 & F250 & F350 & F500 & F850 & F11   \\
    &  (mJy) & (mJy) & (mJy) & (mJy) & (mJy) & (mJy) & (mJy) \\
\hline
1 & \phantom{1}\phantom{1}0.00 $\pm$ 6.91 & \phantom{1}\phantom{1}0.00 $\pm$ \phantom{1}9.54 & \phantom{1}96.68 $\pm$ \phantom{1}6.91 & 133.07 $\pm$ \phantom{1}7.54 & 137.06 $\pm$ \phantom{1}9.84 & 52.7 $\pm$ 0.9 & $\cdots$ \\ 
$.$  & \phantom{1}\phantom{1}0.00 $\pm$ 6.91 & \phantom{1}20.95 $\pm$ \phantom{1}3.34 & \phantom{1}\phantom{1}0.00 $\pm$ 10.36 & \phantom{1}\phantom{1}0.00 $\pm$ \phantom{1}10.00 & \phantom{1}\phantom{1}0.00 $\pm$ 15.43 & 52.7 $\pm$ 0.9 & $\cdots$ \\ 
2 & \phantom{1}\phantom{1}0.20 $\pm$ 2.67 & \phantom{1}12.05 $\pm$ \phantom{1}3.54 & \phantom{1}34.37 $\pm$ \phantom{1}7.11 & \phantom{1}51.07 $\pm$ \phantom{1}7.69 & \phantom{1}41.96 $\pm$ 11.92 & 16.7 $\pm$ 0.9 & $\cdots$ \\ 
3 & \phantom{1}\phantom{1}0.46 $\pm$ 2.25 & \phantom{1}\phantom{1}0.00 $\pm$ \phantom{1}8.72 & \phantom{1}31.71 $\pm$ \phantom{1}6.91 & \phantom{1}27.58 $\pm$ \phantom{1}7.24 & \phantom{1}36.59 $\pm$ \phantom{1}8.47 & 13.0 $\pm$ 0.9 & 3.16 $\pm$ 2.90 \\ 
4 & \phantom{1}\phantom{1}0.00 $\pm$ 6.46 & \phantom{1}\phantom{1}8.41 $\pm$ \phantom{1}3.40 & \phantom{1}10.93 $\pm$ \phantom{1}6.98 & \phantom{1}24.11 $\pm$ \phantom{1}7.69 & \phantom{1}14.10 $\pm$ \phantom{1}8.94 & 11.5 $\pm$ 0.9 & 5.86 $\pm$ 0.51 \\ 
5 & \phantom{1}\phantom{1}4.08 $\pm$ 2.64 & \phantom{1}\phantom{1}6.18 $\pm$ \phantom{1}3.37 & \phantom{1}34.91 $\pm$ \phantom{1}7.07 & \phantom{1}35.78 $\pm$ \phantom{1}7.62 & \phantom{1}31.93 $\pm$ \phantom{1}9.26 & 11.4 $\pm$ 0.9 & $\cdots$ \\ 
\hline
\end{longtable}

\begin{landscape}
\begin{longtable}{rllccccccccccc}
\caption{Optical fluxes of the JCMT/SCUBA2 objects in the COSMOS field. This table is available in its entirety in the online version.}
\label{tab:IDoptCOSMOS}\\
\hline\hline
No. &  RA$_{\rm opt}$ & DEC$_{\rm opt}$ & F0.374 & F0.487 & F0.625 & F0.77 & F0.9 & F1.0 & F1.25 & F1.65 & F2.15 & F3.6 & F4.5  \\
    &  (deg)      & (deg)       & (Jy) & (Jy) & (Jy) & (Jy) & (Jy) & (Jy) & (Jy) & (Jy) & (Jy) & (Jy) & (Jy) \\
\hline
1 & 150.06520 &2.2636520 &4.03e-8$\pm$1.96e-8 & 2.85e-8$\pm$1.79e-8 & 4.65e-8$\pm$2.83e-8 & 8.67e-8$\pm$3.74e-8 & 1.67e-7$\pm$3.84e-8 & 2.48e-7$\pm$1.52e-7 & 1.77e-7$\pm$1.63e-7 & 1.18e-7$\pm$2.30e-7 & 7.89e-7$\pm$3.60e-7 & $\cdots$ & $\cdots$ \\
2 & 149.65820 &2.2356280 &5.67e-8$\pm$1.96e-8 & 4.51e-8$\pm$1.79e-8 & 1.90e-7$\pm$2.83e-8 & 1.60e-7$\pm$3.74e-8 & 9.74e-8$\pm$3.86e-8 & 3.86e-7$\pm$1.52e-7 & 1.09e-6$\pm$1.63e-7 & 1.42e-6$\pm$2.30e-7 & 4.47e-6$\pm$4.47e-7 & 1.20e-5$\pm$1.20e-6 & 1.71e-5$\pm$1.71e-6 \\
3 & $\cdots$ & $\cdots$ &$\cdots$ & $\cdots$ & $\cdots$ & $\cdots$ & $\cdots$ & $\cdots$ & $\cdots$ & $\cdots$ & $\cdots$ & $\cdots$ & $\cdots$ \\
$.$  & 150.03350 &2.4359400 &3.10e-6$\pm$3.10e-7 & 1.15e-5$\pm$1.15e-6 & 4.01e-5$\pm$4.01e-6 & 6.70e-5$\pm$6.70e-6 & 8.94e-5$\pm$8.94e-6 & 0.00$\pm$1.19e-5 & 0.00$\pm$1.57e-5 & 0.00$\pm$2.02e-5 & 0.00$\pm$2.52e-5 & 0.00$\pm$1.90e-5 & 0.00$\pm$1.55e-5 \\
4 & 149.92860 &2.4939160 &5.92e-9$\pm$1.96e-8 & 1.93e-8$\pm$1.79e-8 & 7.51e-8$\pm$2.83e-8 & 3.32e-7$\pm$3.74e-8 & 3.88e-7$\pm$3.88e-8 & 2.96e-7$\pm$1.52e-7 & 3.68e-7$\pm$1.63e-7 & 6.48e-7$\pm$2.30e-7 & 1.64e-6$\pm$3.60e-7 & $\cdots$ & $\cdots$ \\
5 & 150.10010 &2.2971760 &6.69e-7$\pm$6.69e-8 & 3.87e-6$\pm$3.87e-7 & 1.57e-5$\pm$1.57e-6 & 2.74e-5$\pm$2.74e-6 & 3.78e-5$\pm$3.78e-6 & 5.03e-5$\pm$5.03e-6 & 6.80e-5$\pm$6.80e-6 & 8.90e-5$\pm$8.90e-6 & 0.00$\pm$1.16e-5 & 6.64e-5$\pm$6.64e-6 & 5.93e-5$\pm$5.93e-6 \\
\hline
\end{longtable}

\begin{longtable}{rllccccccccccc}
\caption{Optical fluxes of the JCMT/SCUBA2 objects in the UDS field. This table is available in its entirety in the online version.}
\label{tab:IDoptUDS}\\
\hline\hline
No. &  RA$_{\rm opt}$ & DEC$_{\rm opt}$ & F0.374 & F0.487 & F0.625 & F0.77 & F0.9 & F1.0 & F1.25 & F1.65 & F2.15 & F3.6 & F4.5  \\
    &  (deg)      & (deg)       & (Jy) & (Jy) & (Jy) & (Jy) & (Jy) & (Jy) & (Jy) & (Jy) & (Jy) & (Jy) & (Jy) \\
\hline
1 & 34.62774 &-5.5255010 &1.23e-7$\pm$1.23e-8 & 2.31e-7$\pm$2.31e-8 & 3.94e-7$\pm$3.94e-8 & 1.07e-6$\pm$1.07e-7 & 2.71e-6$\pm$2.71e-7 & 5.55e-6$\pm$5.55e-7 & 9.23e-6$\pm$9.23e-7 & 1.54e-5$\pm$1.54e-6 & 2.26e-5$\pm$2.26e-6 & 4.41e-5$\pm$8.82e-6 & 6.06e-5$\pm$1.21e-5 \\
$.$  & $\cdots$ & $\cdots$ &$\cdots$ & $\cdots$ & $\cdots$ & $\cdots$ & $\cdots$ & $\cdots$ & $\cdots$ & $\cdots$ & $\cdots$ & $\cdots$ & $\cdots$ \\
2 & 34.60095 &-5.3825160 &8.40e-8$\pm$1.05e-8 & 6.57e-8$\pm$1.50e-8 & 1.14e-7$\pm$2.00e-8 & 1.47e-7$\pm$2.20e-8 & 1.93e-7$\pm$4.19e-8 & 1.37e-7$\pm$1.39e-7 & 9.01e-8$\pm$7.81e-8 & 3.18e-7$\pm$1.25e-7 & 8.29e-7$\pm$8.29e-8 & 2.96e-6$\pm$5.93e-7 & 4.87e-6$\pm$9.73e-7 \\
3 & 34.83804 &-4.9474300 &3.23e-7$\pm$3.23e-8 & 3.81e-7$\pm$3.81e-8 & 4.79e-7$\pm$4.79e-8 & 5.72e-7$\pm$5.72e-8 & 9.12e-7$\pm$9.12e-8 & 1.07e-6$\pm$1.39e-7 & 1.39e-6$\pm$1.39e-7 & 1.98e-6$\pm$1.98e-7 & 3.69e-6$\pm$3.69e-7 & 7.04e-6$\pm$1.41e-6 & 1.06e-5$\pm$2.11e-6 \\
4 & 34.19973 &-5.0248920 &1.20e-8$\pm$1.05e-8 & 5.77e-8$\pm$1.50e-8 & 1.21e-7$\pm$2.00e-8 & 1.83e-7$\pm$2.20e-8 & 2.33e-7$\pm$4.19e-8 & 2.14e-7$\pm$1.39e-7 & 6.72e-7$\pm$7.81e-8 & 2.06e-6$\pm$2.06e-7 & 5.35e-6$\pm$5.35e-7 & 1.25e-5$\pm$2.51e-6 & 2.09e-5$\pm$4.19e-6 \\
5 & 34.35732 &-5.4282660 &5.28e-6$\pm$5.28e-7 & 9.77e-6$\pm$9.77e-7 & 1.58e-5$\pm$1.58e-6 & 1.95e-5$\pm$1.95e-6 & 2.31e-5$\pm$2.31e-6 & 2.56e-5$\pm$2.56e-6 & 3.14e-5$\pm$3.14e-6 & 4.05e-5$\pm$4.05e-6 & 5.06e-5$\pm$5.06e-6 & 3.72e-5$\pm$7.44e-6 & 4.30e-5$\pm$8.61e-6 \\
\hline
\end{longtable}

\end{landscape}
\begin{longtable}{rccccccccc}
\caption{Redshift and physical properties  of the JCMT/SCUBA2 objects in the COSMOS field. This table is available in its entirety in the online version.}
\label{tab:IDzCOSMOS}\\
\hline\hline
No. &  z$_{\rm opt}$ &  z$_{\rm LW}$ &  SFR      & $M_*$ & frac$_{\rm old}$ & $A_{\rm V,young}$ & $A_{\rm V,old}$ & age$_{\rm young}$ & age$_{\rm old}$\\
    &                &                & (\msunyr) & (\msun) &                 &   (mag)       &   (mag)     & (Gyr)           & (Gyr)  \\
\hline
1 & $0.95_{-0.25}^{+0.45}$ &$3.40_{-0.41}^{+0.57}$ &$828 \pm 47$ &9.18 &0.96 &0.00 &0.00 &0.10 &3.50 \\
2 & $2.08_{-0.08}^{+0.12}$ &$3.10_{-0.23}^{+0.16}$ &$420 \pm 31$ &11.17 &0.08 &4.00 &0.00 &0.20 &0.51 \\
3 & $\cdots$ &$2.70_{-0.13}^{+0.19}$ &$824 \pm 50$ &$\cdots$ &$\cdots$ &$\cdots$ & $\cdots$ & $\cdots$ & $\cdots$ \\
$.$  & $0.37_{-0.12}^{+0.13}$ &$5.40_{-0.52}^{+0.60}$ &$1175 \pm 88$ &10.84 &0.66 &2.00 &0.40 &0.20 &2.00 \\
4 & $4.18_{-0.18}^{+0.27}$ &$3.70_{-0.28}^{+0.27}$ &$1193 \pm 78$ &10.86 &0.68 &0.40 &0.40 &0.10 &1.28 \\
5 & $0.33_{-0.03}^{+0.12}$ &$2.80_{-0.09}^{+0.34}$ &$500 \pm 41$ &10.77 &0.96 &4.00 &0.20 &0.10 &4.00 \\
\hline
\end{longtable}

\begin{longtable}{rccccccccc}
\caption{Redshift and physical properties  of the JCMT/SCUBA2 objects in the UDS field. This table is available in its entirety in the online version.}
\label{tab:IDzUDS}\\
\hline\hline
No. &  z$_{\rm opt}$ &  z$_{\rm LW}$ &  SFR      & $M_*$ & frac$_{\rm old}$ & $A_{\rm V,young}$ & $A_{\rm V,old}$ & age$_{\rm young}$ & age$_{\rm old}$\\
    &                &                & (\msunyr) & (\msun) &                 &   (mag)       &   (mag)     & (Gyr)           & (Gyr)  \\
\hline
1 & $1.40_{-0.05}^{+0.05}$ &$2.50_{-0.05}^{+0.03}$ &$1657 \pm 30$ &11.22 &0.40 &4.00 &0.40 &0.20 &0.72 \\
$.$  & $\cdots$ &$6.00_{-0.22}^{+0.00}$ &$4597 \pm 81$ &$\cdots$ &$\cdots$ &$\cdots$ & $\cdots$ & $\cdots$ & $\cdots$ \\
2 & $3.21_{-0.16}^{+0.09}$ &$2.50_{-0.19}^{+0.11}$ &$1107 \pm 53$ &10.76 &0.97 &0.00 &0.40 &0.09 &1.80 \\
3 & $1.35_{-0.15}^{+0.15}$ &$2.50_{-0.22}^{+0.22}$ &$723 \pm 44$ &10.46 &0.96 &0.00 &0.40 &0.09 &3.50 \\
4 & $3.09_{-0.24}^{+0.16}$ &$2.80_{-0.25}^{+0.42}$ &$650 \pm 37$ &11.54 &0.90 &1.40 &0.40 &0.09 &2.00 \\
5 & $0.45_{-0.10}^{+0.10}$ &$2.30_{-0.17}^{+0.21}$ &$617 \pm 42$ &10.24 &0.93 &0.00 &0.40 &0.09 &1.28 \\
\hline
\end{longtable}

\end{tiny}


\bsp

\label{lastpage}

\end{document}